\documentclass[fleqn,usenatbib]{mnras}

\usepackage{newtxtext,newtxmath}

\usepackage[T1]{fontenc}

\DeclareRobustCommand{\VAN}[3]{#2}
\let\VANthebibliography\thebibliography
\def\thebibliography{\DeclareRobustCommand{\VAN}[3]{##3}\VANthebibliography}

\usepackage{threeparttable}
\usepackage{subdepth}
\usepackage{caption}
\usepackage{multirow}

\usepackage{graphicx}	
\usepackage{amsmath}	

\title[FRB--SRB--XRB]{FRB--SRB--XRB: Geometric and Relativistic Beaming Constraints of Fast Radio Bursts from the Galactic Magnetar SGR J1935+2154}
\author[C.J. Chen and B. Zhang]{
Connery J. Chen,$^{1,2}$\thanks{E-mail: chenc38@unlv.nevada.edu}
Bing Zhang,$^{1,2}$\thanks{E-mail: bing.zhang@unlv.edu}
\\
$^{1}$Nevada Center for Astrophysics, University of Nevada, Las Vegas, NV 89154, USA\\
$^{2}$Department of Physics and Astronomy, University of Nevada, Las Vegas, NV 89154, USA
}

\date{Accepted XXX. Received YYY; in original form ZZZ}

\pubyear{2022}

\begin{document}
\label{firstpage}
\pagerange{\pageref{firstpage}--\pageref{lastpage}}
\maketitle

\begin{abstract}
The detection of a fast radio burst (FRB), FRB 200428, coincident with an X-ray burst (XRB) from the Galactic magnetar soft gamma repeater (SGR) SGR J1935+2154 suggests that magnetars can produce FRBs. Many XRBs have been detected from the source, but very few were found to be associated with bursty radio emission. Meanwhile, a number of weaker radio bursts have been detected from the source, which could in principle be slow radio bursts (SRBs): FRBs detected at viewing angles outside the FRB jet cone.
In this paper, we use these X-ray and radio observations to constrain the geometric and relativistic beaming factors of FRBs under two hypotheses. First, we assume that all SRBs should be associated with XRBs like FRB 200428. We use the FRB--SRB closure relations to identify two SRBs and derive that FRB beaming must be geometrically narrow, $\theta_j \lesssim 10^{-2}$ rad, and follow $\theta_j\Gamma \sim 2$. Second, we assume a less stringent constraint for SRBs by not requiring that they are associated with XRBs. We identify a total of seven SRBs, five of which have Gaussian-like spectra, and derive that FRB beaming factors again follow $\theta_j\Gamma \sim 2$.
\end{abstract}




\section{Introduction}
On 2020 April 28, a 1.5 MJy fast radio burst (FRB) originating from the Galactic magnetar SGR J1935+2154 was detected by the Suvery for Transient Astronomical Radio Emission (STARE2) radio array and Canadian Hydrogen Intensity Mapping Experiment (CHIME) at equal arrival times \citep{bochenek_frb_2020, chimefrb_collaboration_frb_2020}. This event is referred to as FRB 200428. Remarkably, a hard X-ray burst (XRB) was independently detected by Insight--HXMT, INTEGRAL, AGILE, and Konus-Wind \citep{li_hxmt_2021, mereghetti_integral_2020, tavani_agile_2021, ridnaia_kw_2021}. It was the first associated FRB--XRB event ever detected; whether it was an intrinsically unique event from the majority of FRB-absent SGR bursts is still subject to debate \citep[i.e.,][]{yang_bursts_2021}. Extensive monitoring of the source in radio and X-ray bands has enabled investigations on whether the observed scarcity might be due to beaming \citep{lin_no_2020,scholz_simultaneous_2017}.

The Five-hundred-meter Aperture Spherical Telescope (FAST) first observed SGR J1935+2154 on 2020 April 15 UTC 21:54:29 and began monitoring the source daily on 2020 April 26 UTC 21:06:55. FAST was not monitoring SGR J1935+2154 at the time of FRB 200428. On 2020 April 30 UTC 21:43:00.42, FAST detected one weak and highly polarized radio emission from the source \citep{2020ATel13699....1Z}. During the observation windows conducted by FAST in April, the Fermi Gamma-ray Burst Monitor (Fermi/GBM) reported 34 XRBs from SGR J1935+2154 \citep{yang_bursts_2021} and the Neutron star Interior Composition Explorer (NICER) reported 121, 104 of which are unique from the Fermi/GBM detections \citep{younes_nicer_2020}. FAST continued to monitor the source almost daily until May 19 but detected no additional radio signals \citep{zhu_birth_submitted}. During these extended observations, Insight--HXMT detected two XRBs from the source on May 6 and May 12 \citep{2022ApJS..260...24C}. Numerous non-detections of radio emission by FAST during SGR bursting events indicate that FRB--XRB associations are exceptionally rare.

Targeted observations in the following months of 2020 by the Westerbork Radio Telescope 1 (RT1), the Northern Cross at the Medicina Radio Astronomical Station (Medicina Northern Cross; MNC), CHIME, FAST, and the Big Scanning Array of Lebedev Physical Institute (BSA/LPI) yielded several additional weak radio bursts \citep{kirsten_detection_2021, 2020ATel13783....1B, 2020ATel14074....1G, 2020ATel14084....1Z, 2020ATel14186....1A}. None of the radio detections following FRB 200428 were found to be associated with an XRB.

In early October, 2022, SGR J1935+2154 entered another active bursting phase \citep{2022ATel15667....1P}. At least two additional weak radio bursts were detected from the source \citep{2022ATel15681....1D, 2022ATel15707....1H}, both accompanied by simultaneous XRB detections \citep{2022ATel15682....1W, 2022ATel15708....1L}.

In Section~\ref{sec:frb_srb} we reiterate the theory of SRBs introduced in \cite{zhang_slow_2021}, extend the application to FRBs with a Gaussian spectrum and discuss FRB beaming constraints by SRB detectability. In Section~\ref{sec:frb_xrb} we review FRB beaming constraints by non-detections of radio emission by FAST concurrent with XRBs \citep[e.g.,][]{lin_no_2020} and extend the models to FRBs with a Gaussian spectrum. A collection of observational data is presented in Section~\ref{sec:observations}. The theoretical models are applied to observational data under two hypotheses, discussed in Section~\ref{sec:results}. The conclusions are presented in Section~\ref{sec:conclusions}.

\section{On-beam FRB vs. Off-beam SRB}
\label{sec:frb_srb}

The original definition and description of SRBs is presented in \cite{zhang_slow_2021}. SRB theory considers geometric and relativistic beaming factors for FRBs. Current FRB emission models postulate that relativistically moving plasma produces the FRB emission \citep[][and references therein]{zhang_frb_2020}. When such emission is beamed (non-isotropic), SRBs should be detectable. 

Observations suggest that at least some FRBs, especially repeating sources, have intrinsically narrow spectra consistent with a Gaussian shape \citep[e.g.,][]{aggarwal_frb121102_2021,2020ATel14080....1P}. In this work we extend the theories to apply to FRBs with narrow Gaussian spectra.

\subsection{Doppler factor}

Consider a relativistic conical jet with bulk Lorentz factor $\Gamma$ (corresponding to dimensionless speed $\beta$) and half opening angle $\theta_j$. Also consider an observer viewing at some angle $\theta$ from the jet axis. The Doppler Factor can be defined as
\begin{equation}
\label{eq:doppf}
    \mathcal{D}(\theta) = 
    \begin{cases}
        \mathcal{D}_{\text{on}} = \frac{1}{\Gamma(1 - \beta)}, & \theta < \theta_j \\
        \mathcal{D}_{\text{off}} = \frac{1}{\Gamma(1-\beta \cos{[\Delta \theta)}]}, & \theta > \theta_j
    \end{cases}
\end{equation}
where $\Delta \theta = \theta-\theta_j$ \citep{zhang_slow_2021}.  In a relativistic context, the Doppler factor connects observer-frame (unprimed) quantities with comoving-frame (primed) quantities (e.g., \cite{zhang_grb_2018})
\begin{align}
    \Delta t &=\mathcal{D}^{-1}\Delta t' \\
    \nu &= \mathcal{D}\nu' \\
    L_{\nu}(\nu) &= \mathcal{D}^3L'_{\nu'}(\nu')
\end{align}
where $\Delta t$ is the emission duration, $\nu$ is the emission frequency, and $L_{\nu}$ is the isotropic-equivalent specific luminosity. The last expression assumes a point source for which all emitter material is moving towards one direction. This is justified for FRBs because the emission region is very small.

\begin{figure}
    \centering
    \includegraphics[scale=0.23]{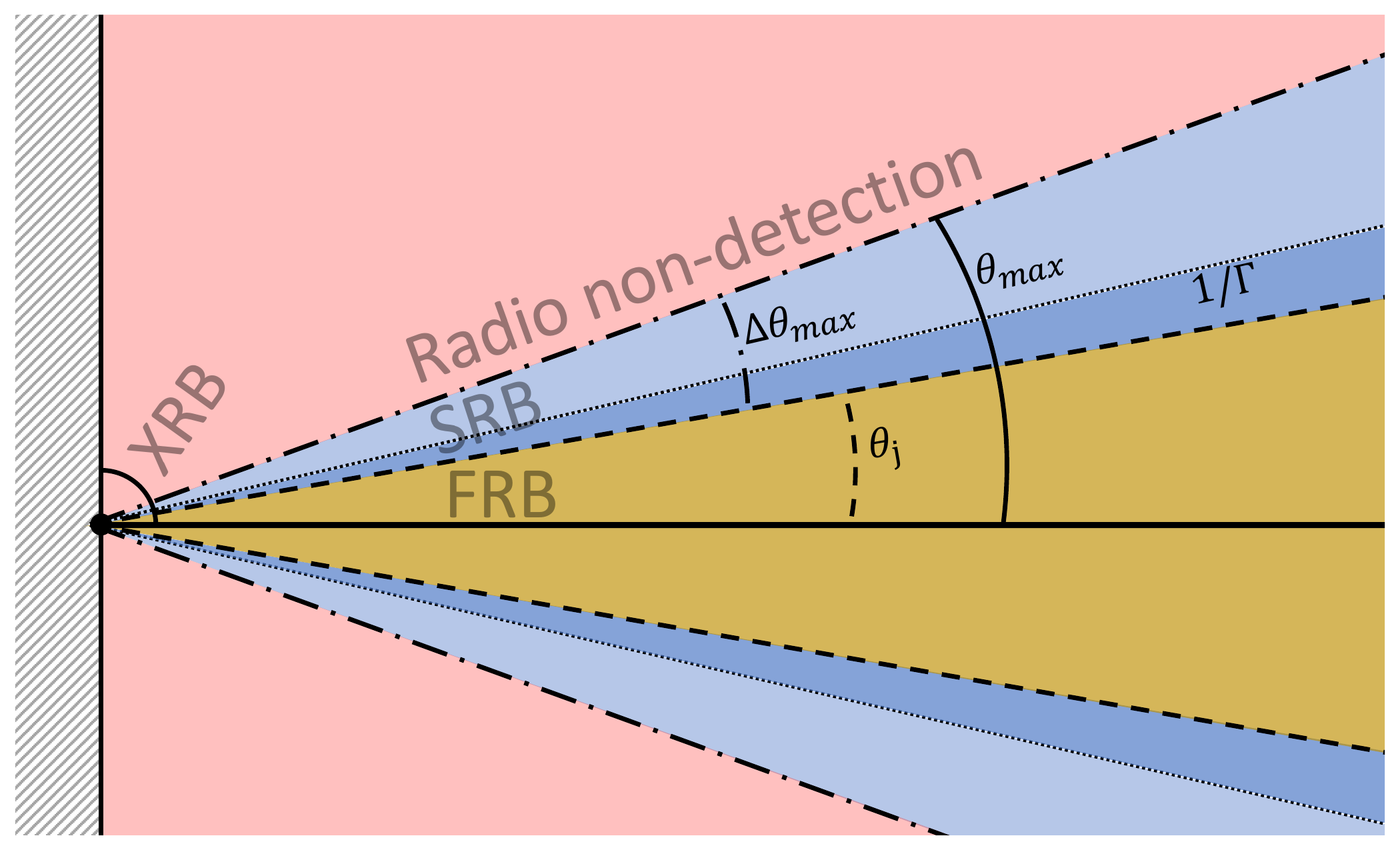}
    \caption{Geometry of the FRB--SRB--XRB beaming interpretation. A beamed FRB (\textit{gold, dashed}) with half-opening angle $\theta_j$ is detectable as an off-axis SRB to a certain maximum angle beyond the jet cone $\Delta\theta_{\text{max}}$ (\textit{blues, dash-dotted}), corresponding to a maximum viewing angle from the jet axis, $\theta_{\text{max}}=\Delta\theta_{\text{max}}+\theta_j$ (\textit{black, solid}). A viewing angle beyond $1/\Gamma$ (dark-blue, dotted) will witness a rapid decline in SRB fluence. Beyond $\theta_{\text{max}}$, in the red region, an observer will not detect any radio signal, but instead will only detect the associated XRB emitted across 2$\pi$ steradians. The radio emission region is, at maximum, 2$\pi$ steradians, (in this geometric representation, $\theta_{\text{max}} \leq \pi/2$). The hatched region behind the emission site is beyond the possible region of emission in this orientation. The magnetar source is assumed to be a point source, represented by the black dot.}
    \label{fig:doppf_geometry}
\end{figure}

Consider two observers, one viewing the jet at an on-axis angle $\theta \leq \theta_j$ so that $\mathcal{D} = \mathcal{D}_\text{on}$, and the other viewing the jet at an off-axis angle $\theta \geq \theta_j$ so that $\mathcal{D} = \mathcal{D}_\text{off}$. As in \cite{zhang_slow_2021}, we compare the observed properties by defining a ratio of Doppler factors
\begin{equation}
\label{eq:r_d}
    \mathcal{R}_{\mathcal{D}} \equiv \frac{\mathcal{D}_{\text{on}}}{\mathcal{D}_{\text{off}}} \geq 1.
\end{equation}
Assuming that the comoving-frame parameters are identical for both on- and off-axis observers, $\mathcal{R}_{\mathcal{D}}$ directly relates the on-beam FRB properties with the off-beam SRB properties. The general geometry of the beaming interpretation is shown in Fig.~\ref{fig:doppf_geometry}.

In this work we only consider a narrow Gaussian-like spectrum for FRBs in the rest-frame. The specific luminosity spectrum  \citep[e.g.,][]{zhang_slow_2021} is
\begin{equation}
\label{eq:specific_luminosity_gaussian}
    L'_{\nu'}(\nu')= L'_{\nu'}(\nu_0')\exp{\left[-\frac{1}{2}\left(\frac{\nu'-\nu_0'}{\delta\nu'}\right)^2\right]}
\end{equation}
where $\nu'_0$ and $\delta\nu'$ are the central frequency and characteristic width of the spectrum, respectively. In the case of a narrow spectrum, $\delta\nu'/\nu' \ll 1$.

\subsection{FRB--SRB closure relation}
\label{sec:frb_srb_closure}

Consider again one on-axis FRB observer and one off-axis SRB observer. For an FRB (SRB) with specific fluence $\mathcal{F}_{\nu}^{\text{FRB}}$ ($\mathcal{F}_{\nu}^{\text{SRB}}$), intrinsic width $w^{\text{FRB}}$ ($w^{\text{SRB}}$), and observing frequency at $\nu_1$ ($\nu_2$), one may assume $\nu_1=\mathcal{D}_{\text{on}}\nu'_0=\nu^{\text{FRB}}$ ($\nu_2=\mathcal{D}_{\text{off}}\nu'_0=\nu^{\text{SRB}}$). Using equation~\ref{eq:specific_luminosity_gaussian} together with the above, the ratio of specific luminosities can be written as 

\begin{align}
\label{eq:luminosity_ratio}
    \frac{L_{\nu}^{\text{off}}(\nu_2)}{L_{\nu}^{\text{on}}(\nu_1)} &=
    \frac{\mathcal{D}^{3}_{\text{off}}}{\mathcal{D}^{3}_{\text{on}}}
    \frac{\exp\left[-\frac{1}{2}\left(\frac{\nu_{2}-\nu^{\text{off}}}{\delta\nu^{\text{off}}}\right)^2\right]}{\exp\left[-\frac{1}{2}\left(\frac{\nu_{1}-\nu^{\text{on}}}{\delta\nu^{\text{on}}}\right)^2\right]} \notag \\
    &= \mathcal{R}_\mathcal{D}^{-3} \exp\left\{-\frac{1}{2}\left[\left(\frac{\nu'_2-\nu'_0}{\delta\nu'}\right)^2-\left(\frac{\nu'_1-\nu'_0}{\delta\nu'}\right)^2\right]\right\}
\end{align}
\citep{zhang_slow_2021}. For an on-axis FRB observer, one may assume that the observing frequency is at the central frequency of the spectrum, i.e., $\nu'_0=\nu'_1$. Simplifying equation~\ref{eq:luminosity_ratio}, one arrives at an expression for the luminosity ratio between the off-axis SRB and on-axis FRB observers.
\begin{align}
\label{eq:luminosity_ratio_simpl}
    \frac{L_{\nu}^{\text{off}}(\nu_2)}{L_{\nu}^{\text{on}}(\nu_1)} &= \mathcal{R}_\mathcal{D}^{-3} \exp\left\{-\frac{1}{2}\left[\left(\frac{\nu'_2-\nu'_1}{\delta\nu'}\right)^2\right]\right\} \notag \\
    & = \mathcal{R}_\mathcal{D}^{-3} \exp\left[-\frac{1}{2}\left(\frac{\frac{\nu_2}{\mathcal{D}_{\text{off}}}-\frac{\nu_1}{\mathcal{D}_{\text{on}}}}{\frac{\delta\nu_{\text{on}}}{\mathcal{D}_{\text{on}}}}\right)^2\right] \notag \\
    &= \mathcal{R}_\mathcal{D}^{-3} \exp\left[-\frac{1}{2}\left(\frac{\nu_1}{\delta\nu_{\text{on}}}\right)^2\left(\mathcal{R}_{\mathcal{D}}\frac{\nu_2}{\nu_1}-1\right)^2\right].
\end{align}
Noticing that $\mathcal{R}_{\mathcal{D}} = w^{\text{off}}/w^{\text{on}}$, \cite{zhang_slow_2021} obtains a closure relation for this ratio 
\begin{equation}
\label{eq:frb_srb_closure_gaussian}
    \left(\frac{F_{\nu}^{\text{SRB}}}{F_{\nu}^{\text{FRB}}}\right)
    \left(\frac{w_{\nu}^{\text{SRB}}}{w_{\nu}^{\text{FRB}}}\right)^2
    \exp
    \left[\frac{1}{2}
        \left(
            \frac{\nu^{\text{SRB}}}{\delta\nu^{\text{FRB}}}
        \right)^2
        \left
            (\frac{w^{\text{SRB}}}{w^{\text{FRB}}}-1
        \right)^2
    \right]
    = 1.
\end{equation}
For data sets that span many orders of magnitude, as is usually the case when dealing with FRBs and SRBs, it is more feasible to consider a logarithmic form of the closure relation, i.e.,
\begin{equation}
\label{eq:frb_srb_closure_gaussian_log}
    \ln\left({\frac{F_{\nu}^{\text{SRB}}}{F_{\nu}^{\text{FRB}}}}\right)
    + 2\ln\left({\frac{w_{\nu}^{\text{SRB}}}{w_{\nu}^{\text{FRB}}}}\right) 
    + \left[
        \frac{1}{2}
        \left(
            \frac{\nu^{\text{SRB}}}{\delta\nu^{\text{FRB}}}
        \right)^2
        \left
            (\frac{w^{\text{SRB}}}{w^{\text{FRB}}}-1
        \right)^2
    \right]
    = 0.
\end{equation}
Equations~\ref{eq:frb_srb_closure_gaussian} or \ref{eq:frb_srb_closure_gaussian_log} can be used to determine whether observed radio pulses are off-axis SRB viewings of on-axis FRBs. In case we must use equation~\ref{eq:frb_srb_closure_gaussian_log} due to computational limits, we confirm unity by equating it to the parameter, $u=\log(1+\epsilon)\approx0$ where $\epsilon\ll 1$.

\subsection{SRB Detectability}
\label{sec:detectability}

\begin{figure*}
    \centering
    \includegraphics[scale=0.33]{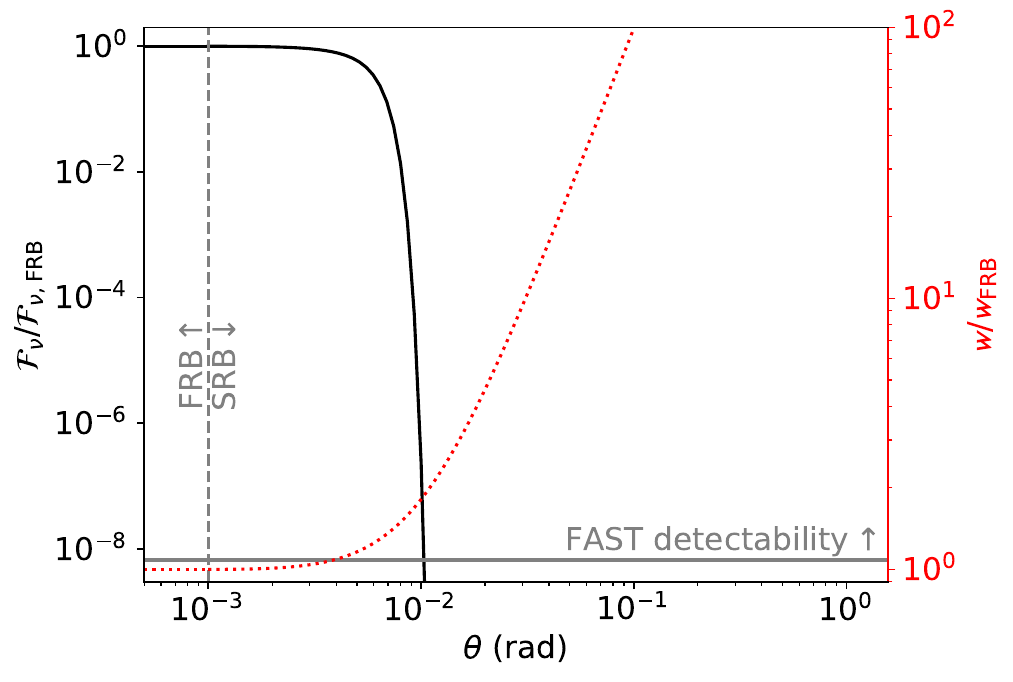}
    \includegraphics[scale=0.33]{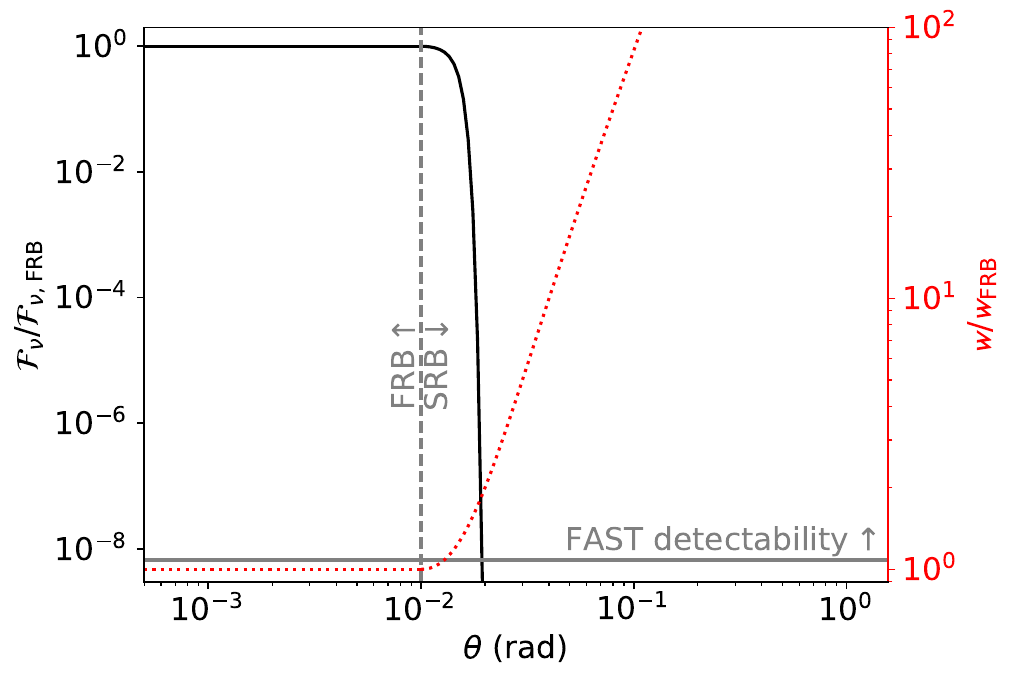}
    \includegraphics[scale=0.33]{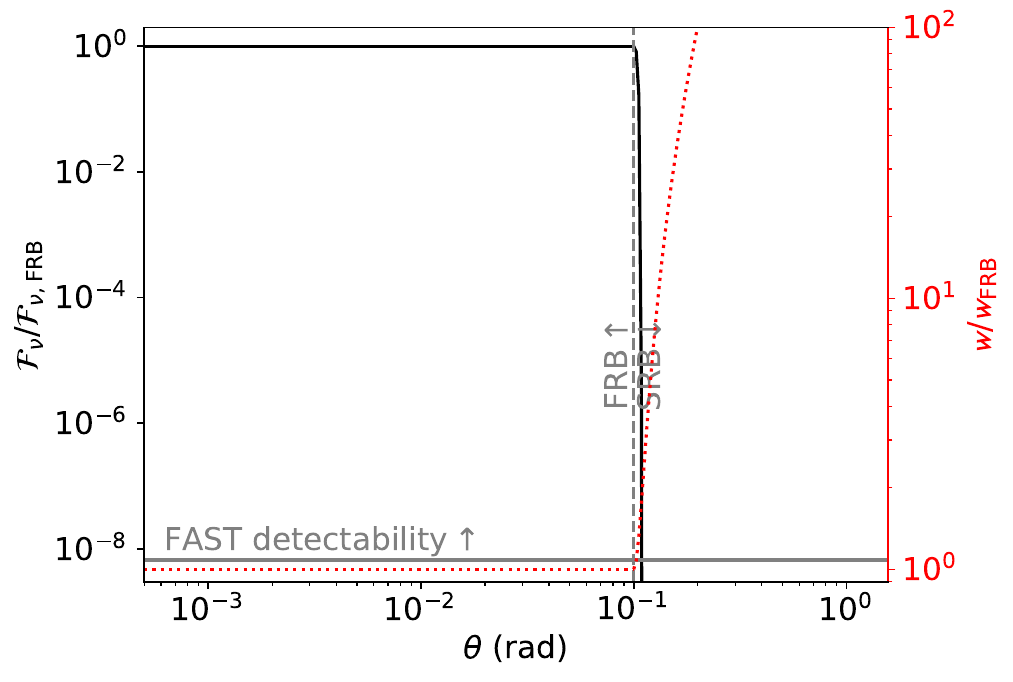}
    \caption{ Burst specific fluence ({\it left axis}) and pulse duration ({\it right axis}) expressed in terms of on-beam FRB properties ($\mathcal{F_\text{FRB}}=1.5$ MJy ms and $w_{\text{FRB}}=0.61$ ms for FRB 200428) as a function of viewing angle. The vertical line ({\it grey, dashed}) represents the presumed geometric width of the FRB beam, $\theta_j$. In each panel, $\theta_j$ is varied ({\it left:} $\theta_j=10^{-3}$ rad, {\it middle:} $\theta_j=10^{-2}$ rad, {\it right:} $\theta_j=10^{-1}$ rad) while $\Gamma=100$ is unchanged. Notice in the left panel that when $\theta_j<1/\Gamma$, the FRB specific fluence is mostly preserved until $\theta\sim1/\Gamma$, and conversely in the right panel that when $\theta_j>1/\Gamma$, the FRB specific fluence drops off very sharply beyond $\theta_j$. The horizontal line ({\it grey, solid}) represents the FAST sensitivity threshold, above which radio emission is detectable by FAST.}
    \label{fig:srb_theta}
\end{figure*}

Beyond the FRB jet cone, $\theta_j$, the specific fluence of an off-axis SRB diminishes rapidly and the pulse duration increases. The variation depends on the intrinsic beaming factors of the FRB. FRB emission models invoke a highly relativistic plasma, $\Gamma\gtrsim10^2$ \citep{lyubarsky_08,murase_16,lu_kumar_16,Zhang_ICS_22} to produce the FRB emission, and may be narrowly beamed along open magnetic field lines if the emission region is within the magnetosphere \citep{Kumar17, Yang_18, metzger_19, Wadiasingh_2019, wang_magnetospheric_2020, kumar_20, lu_unified_2020, yang_pair_20, Wadiasingh_2020, Qu21, Yang_21, Wang_22, Zhang_ICS_22}, see the comprehensive review by \cite{zhang_RMP_22}. These theoretically motivated suggestions are also supported by the results of this work, as will be discussed in Section~\ref{sec:results}. With two independent detections of FRB 200428 in different wavelength bands by CHIME and STARE2, we find that the spectrum can be fitted as a Gaussian with a narrow characteristic spectral width referenced to the central frequency, $\delta\nu/\nu_0 = 0.278$ (Appendix~\ref{appx:spectral_width}). Fig.~\ref{fig:srb_theta} demonstrates how specific fluence and pulse duration of FRB 200428 change beyond the FRB jet cone, assuming typical values of $\Gamma$ and $\theta_j$ according to the above discussion.

The viewing angle, $\theta$, must not lie too far outside the FRB jet cone so the specific fluence does not drop below the telescope sensitivity threshold. There exists a maximum viewing angle, $\theta_{\text{max}}$, that defines a maximum FRB jet offset angle $\Delta \theta_{\text{max}}=\theta_{\text{max}}-\theta_j$ (see Fig.~\ref{fig:doppf_geometry}). That is, $\Delta \theta_{\text{max}}$ is the viewing angle beyond the jet cone at which the SRB specific fluence diminishes to the telescope sensitivity threshold, $\mathcal{F}^{\text{SRB}}=\mathcal{F}_{\text{th}}$.

Combining equations~\ref{eq:doppf}, \ref{eq:r_d} and \ref{eq:frb_srb_closure_gaussian}, one gets
\begin{equation}
\label{eq:dtheta_max}
    \Delta\theta_{\max} = \cos^{-1}\left[\frac{1-\mathcal{R}_{\mathcal{D}}(1-\beta)}{\beta}\right]
\end{equation}
where $\mathcal{R}_{\mathcal{D}} = w^{\text{SRB}}/w^{\text{FRB}}$ is obtained by numerically solving equation~\ref{eq:frb_srb_closure_gaussian} or \ref{eq:frb_srb_closure_gaussian_log} with $\mathcal{F}^{\text{SRB}}=\mathcal{F}_{\text{th}}$.

\begin{figure*}
    \centering
    \includegraphics[scale=0.5]{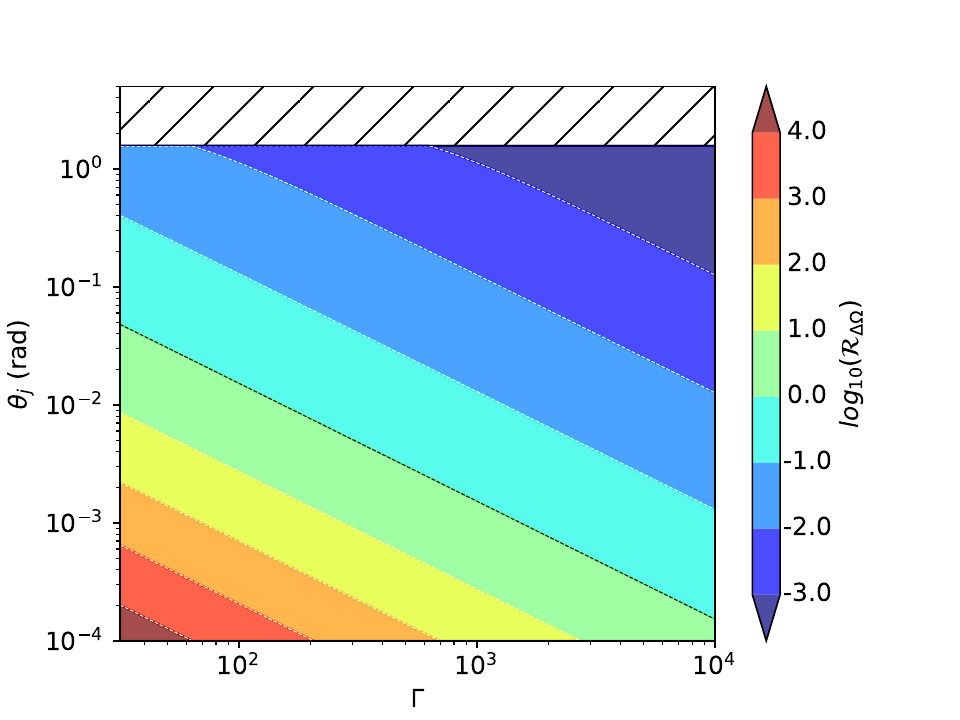}
    \includegraphics[scale=0.5]{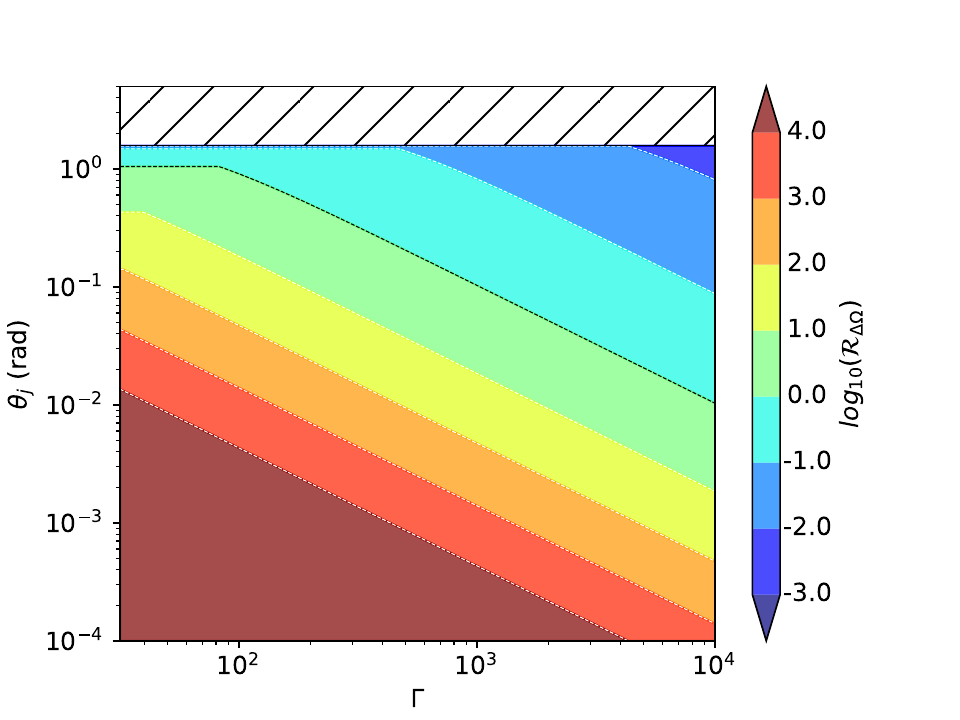}
    \caption{FRB beaming constraints for FRB 200428. The left panel shows the case of a narrow spectrum, $\delta\nu/\nu=0.1$; the right, a wide (i.e., flat) spectrum, $\delta\nu/\nu=10^3$. The black dotted line represents $\Omega_{\text{SRB}} = \Omega_{\text{FRB}}$, or $\mathcal{R}_{\Delta\Omega} = 1$. Constant constraints for the geometric FRB beaming angle, especially apparent FRBs with flat spectra, is due to the off-axis SRBs being visible at all viewing angles, $\theta_{\text{max}} = \pi/2$. The hatched region represents the geometrically forbidden and unphysical region of FRB beaming, $\theta_j>\pi/2$.}
    \label{fig:frb_srb_compare}
\end{figure*}

In this beaming interpretation, the half opening angle of radio emission, $\theta_{\text{max}}=\Delta\theta_{\text{max}}+\theta_j$, has a maximum value of $\pi/2$. If the FRB jet is geometrically wide ($\theta_j \sim \pi/2$) and/or not highly relativistic ($\Gamma \sim 1$), then radio emission is detectable at all viewing angles, $\theta_{\max} = \pi/2$. The effect is apparent in the upper-left region of the right panel of Fig.~\ref{fig:frb_srb_compare}, where $\theta_j$ is large and $\Gamma$ is small.

\cite{zhang_slow_2021} defines a solid angle ratio between detectable Galactic SRBs and FRBs,
\begin{equation}
\mathcal{R}_{\Delta\Omega} \equiv \frac{\Delta\Omega^{\text{SRB}}}{\Delta\Omega^{\text{FRB}}}
\end{equation} 
where $\Delta\Omega^{\text{SRB}}$ and $\Delta\Omega^{\text{FRB}}$ are the solid angles of the SRB and FRB jet beam, respectively. Expanding and simplifying $\Delta\Omega^{\text{SRB}}$ and $\Delta\Omega^{\text{FRB}}$,
\begin{align}
\label{eq:r_sa}
    \mathcal{R}_{\Delta\Omega} &=\frac{\int_0^{2\pi}\int_{\theta_j}^{\Delta\theta_{\text{max}}+\theta_j} \sin\theta d\theta d\phi}{\int_0^{2\pi}\int_{0}^{\theta_j} \sin\theta d\theta d\phi} \notag\\
    &= \frac{\cos\theta_j-\cos(\Delta\theta_{\text{max}}+\theta_j)}{1-\cos\theta_j}.
\end{align}
In the case where the SRB is detectable at all viewing angles beyond the FRB jet cone, $\Delta\theta_{\text{max}}+\theta_j = \pi/2$, equation~\ref{eq:r_sa} simplifies to
\begin{equation}
    \mathcal{R}_{\Delta\Omega}= \frac{\cos\theta_j}{1-\cos\theta_j}.
\end{equation}
Equation~\ref{eq:r_sa} represents the ratio of the detectable region of an SRB to that of an FRB, given certain FRB beaming properties and a particular telescope sensitivity threshold. In this work we consider FAST with a fluence sensitivity threshold of $\mathcal{F}_{\text{th, FAST}} \approx 10 \text{ mJy ms}$ for a 1 ms burst \citep{lin_no_2020, fast_collaboration_commissioning_2019}.

The solid angle ratio, $\mathcal{R}_{\Delta\Omega}$, is analogous to the ratio of detected SRBs to FRBs from Galactic magnetars. Contours of geometric and relativistic FRB beaming factors for various solid angle ratios are shown in Fig.~\ref{fig:frb_srb_compare} for a narrow FRB spectrum (\textit{left}: $\delta\nu/\nu\sim0.1$) and a wide, essentially flat FRB spectrum (\textit{right}: $\delta\nu/\nu\sim10^3$), where $\delta\nu/\nu$ is the characteristic spectral width of the FRB referenced to the emission frequency. Notice that for any given $\mathcal{R}_{\Delta\Omega}$, FRB beaming is more constrained when the spectrum is narrow versus when it is wide. This suggests that an FRB with a narrow Gaussian spectrum is more sensitive to relativistic beaming effects, i.e., it will undergo a more rapid decline in fluence beyond the FRB beam.

\section{Radio Non-Detections during SGR Bursting Phases}
\label{sec:frb_xrb}

Since FRB 200428 was associated with an XRB, we also consider FRB beaming constraints based on non-detections of radio emission concurrent with XRBs. FAST is presently the world's most sensitive radio telescope (see Section~\ref{sec:fast}), and its extensive monitoring of SGR J1935+2154 provides substantial data to constrain the FRB/SRB--XRB association.

The flux contrast of non-detections should be
\begin{equation}
\label{eq:frb_fast_ratio}
    \frac{F_{\nu}^{\text{FRB}}}{F_{\nu}^{\text{th, FAST}}} =
    \frac{f_{\nu}^{\text{FRB}}}{f_{\nu}^{\text{th, FAST}}} \gtrsim \eta = 1.5\times10^8
\end{equation}
where $f_{\nu}^{\text{FRB}}\gtrsim1.5 \text{ MJy ms}$ is the fluence of FRB 200428, and $f_{\nu}^{\text{th, FAST}}\approx 10 \text{ mJy ms}$ for a 1 ms burst is the fluence sensitivity threshold of FAST. $F_{\nu}^{\text{FRB}}$ and $F_{\nu}^{\text{th, FAST}}$ are the corresponding flux densities \citep{lin_no_2020}.

\begin{figure*}
    \centering
    \includegraphics[scale=0.5]{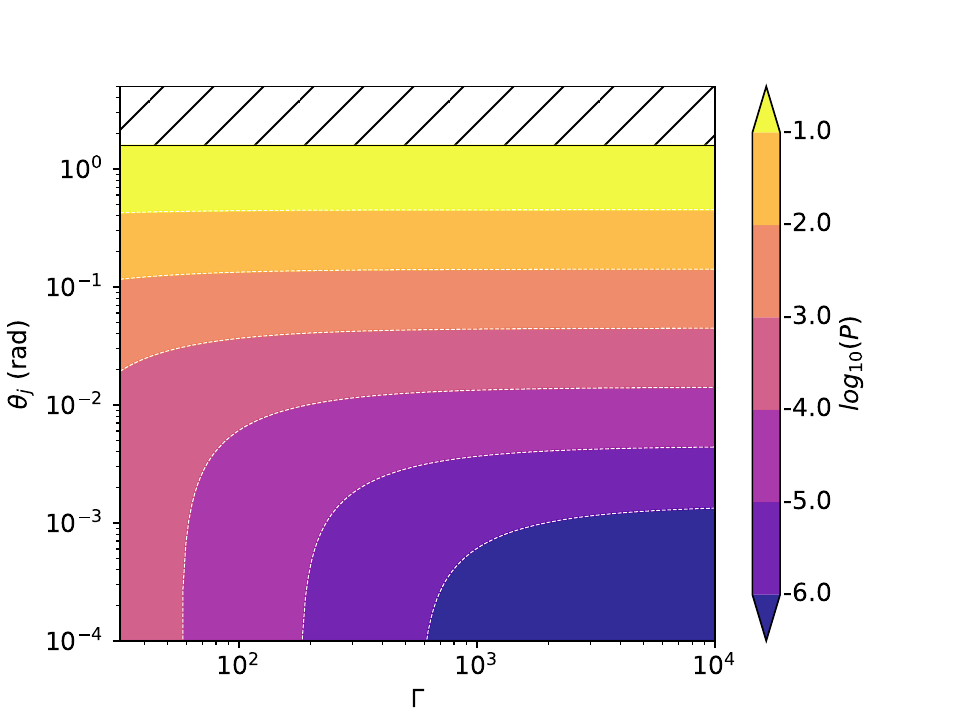}
    \includegraphics[scale=0.5]{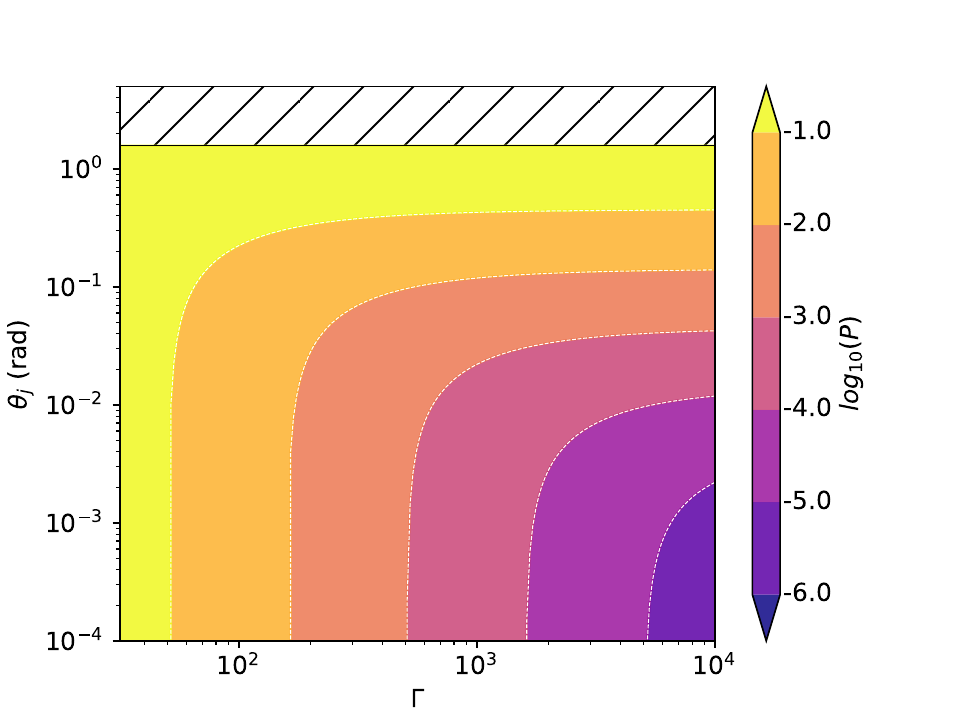}
    \caption{FRB beaming constraints by non-detections of radio emission by FAST concurrent with XRBs for a FRB 200428 reference FRB assuming isotropic XRB emission. The left panel shows the case of a narrow FRB spectrum, $\delta\nu/\nu=0.1$; the right, a wide (i.e., flat) FRB spectrum, $\delta\nu/\nu=10^3$. The Lorentz factor is more difficult to constrain for FRBs with narrow Gaussian spectra (\textit{left}) compared to FRBs with flat spectra (\textit{right}). This is because the former is more sensitive to relativistic beaming effects compared to the latter. Notice that geometric beaming is unaffected by the spectral width of the Gaussian FRB. The hatched region represents the forbidden unphysical region of geometric FRB beaming, $\theta_j>\pi/2$.}
    \label{fig:frb_xrb_compare}
\end{figure*}

Consider the flux ratio, $\eta$, which is identical to the luminosity ratio given the same source, equation~\ref{eq:luminosity_ratio_simpl}. We assume that FRB 200428 was viewed on-axis, $\theta < \theta_j$, and that XRB-associated FRBs not detected by FAST have $\theta > \theta_j$, and so
\begin{multline}
\label{eq:frb_xrb}
    \left[\frac{1-\beta\cos(\Delta\theta)}{1-\beta}\right]^3\\
    \times \exp\left\{\frac{1}{2}\left(\frac{\nu_1}{\delta\nu^{\text{on}}}\right)^2\left(\frac{1-\beta\cos(\Delta\theta)}{1-\beta}\frac{\nu_2}{\nu_1}-1\right)^2\right\} \gtrsim \eta.
\end{multline}
Equation~\ref{eq:frb_xrb} can be numerically solved as an equality to obtain a characteristic viewing angle, $\Delta\theta_c$, so that the above condition will be satisfied if $\Delta\theta > \Delta\theta_c$. We assume that SGR burst emission is $2\pi$ steradians\footnote{\cite{lin_no_2020} previously provided FRB beaming constraints in this fashion but assumed that the SGR burst emission is isotropic ($4\pi$ steradians) and that the FRB spectrum is flat (see Extended Data Fig. 4 therein, which is consistent with Fig.~\ref{fig:frb_xrb_compare} right panel).} so that the probability for FRB/SRB--XRB associations is
\begin{equation}
\label{eq:p_0}
    P_0 \geq P \simeq \frac{1}{2\pi}\left(2\pi\int_0^{\theta_j+\Delta\theta_c}\sin\theta d\theta \right) = 1-\cos(\theta_j+\Delta\theta_{c}).
\end{equation}

The true probability of FAST detecting an SRB associated with an XRB, $P$, is probably less than the observed ratio of SRB-associated to SRB-absent XRBs, $P_0$. Therefore, 
\begin{equation}
    \theta_j \leq \arccos(1-P_0)-\Delta\theta_c.
\end{equation}
Contours of geometric and relativistic FRB beaming factors for various detection probabilities are shown in Fig.~\ref{fig:frb_xrb_compare}. 
In the left panel, the FRB spectrum is assumed to be narrow ($\delta\nu/\nu\sim0.1$), in the right panel, wide ($\delta\nu/\nu\sim10^3$). The Lorentz factor is more difficult to constrain for an FRB with a narrow Gaussian spectrum compared to an FRB with a flat spectrum because the former is more sensitive to relativistic beaming effects compared to the latter. That is, a highly relativistic FRB with a flat spectrum has an off-axis SRB comparable to that of a less relativistic FRB with a narrower spectrum.

\section{Observational Data}
\label{sec:observations}

This work incorporates a comprehensive collection of X-ray and radio observations of SGR J1935+2154 conducted by various telescopes in order to obtain well-sampled constraints to FRB beaming. Section~\ref{observations_xray} presents the X-ray observations relevant to this work, namely XRBs detected with associated radio emission and XRBs detected during FAST observing windows wherein no pulsed radio emission was detected. Section~\ref{sec:observations_radio} presents eight radio bursts detected from the source following FRB 200428 and outlines the radio observation epochs conducted by FAST concurrent with 140 XRB detections not associated with pulsed radio emission.

\subsection{X-ray observations}
\label{observations_xray}

On 2020 April 27 UTC 18:26:20, the Swift Burst Alert Telescope (Swift/BAT) and Fermi/GBM were triggered by an active phase of the Galactic magnetar SGR J1935+2154 \citep{2020GCN.27665....1S,2020GCN.27659....1F}. Shortly thereafter, a series of X-ray telescopes began target-of-opportunity observations of the source. Throughout the 2020 active bursting phase, only one XRB was detected in association with radio emission (FRB 200428), while 140 XRBs were detected during FAST observing windows wherein no pulsed radio emission was detected. 

On 2022 October 10, Swift/BAT and INTEGRAL detected a number of XRBs from SGR J1935+2154 \citep{2022ATel15667....1P,2022GCN.32698....1M}, indicating that the magnetar was entering another active bursting phase. In the following weeks, two XRBs were detected \citep{2022ATel15682....1W, 2022ATel15708....1L}, both with arrival times coincident with the detection of radio bursts \citep{2022ATel15681....1D, 2022ATel15707....1H}.

The various X-ray observation campaigns of the source are outlined in this section, and the relevant XRBs are listed in Appendix~\ref{appx:xray_observations}.

\subsubsection{Fermi/GBM}
The Fermi Gamma-ray Space Telescope was launched on 2008 June 11 carrying two scientific instruments, the Large Area Telescope (LAT) and the GBM. The LAT observes gamma-rays above $\sim$20 MeV while the GBM observes gamma-rays between $\sim$8 keV and $\sim$40 MeV \citep{2009ApJ...702..791M}.

On 2020 27 April at UTC 18:26:20.16, Fermi/GBM triggered on a bright, SGR-like burst from the direction of the magnetar SGR J1935+2154. The burst had a $T_{90}$ duration of $\sim$2 seconds with a fluence of $20 \pm 2 \times 10^{-8}$ erg cm$^{-2}$ s$^{-1}$ in the energy range of 10--200 keV. Fermi/GBM subsequently triggered on multiple bursts from SGR 1935+2154, indicating that the magnetar was entering an active bursting phase \citep{2020GCN.27659....1F}.

\cite{lin_no_2020} reported no pulsed radio emission from SGR J1935+2154 detected in association with 29 XRBs detected by Fermi/GBM \citep[see also][]{zou_periodicity_2021}. More recently, \cite{yang_bursts_2021} performed a thorough untriggered burst search on Fermi/GBM data to identify 34 radio-absent XRBs during the same FAST observing session between 2020 April 27 UTC 23:55:00 and 2020 28 April 00:50:37. Continuous Fermi/GBM data from 2013 January to 2021 October reveals no additional XRB detections during FAST observing sessions through 2020 October \citep{lin_fermigbm_2020, zhu_birth_submitted}. In this work we include 34 XRBs from SGR J1935+2154 detected by Fermi/GBM during simultaneous monitoring by FAST, based on the more recent untriggered burst search in \cite{yang_bursts_2021}.

\subsubsection{Insight--HXMT}

The Insight Hard X-ray Modulation Telescope (HXMT) is China's first astronomical satellite launched on 2017 June 15. It contains three detectors for low-, medium- and high- energy emission with a total energy range of 1--250 keV. \citep{2018SPIE10699E..1UZ}.

After Swift/BAT and Fermi/GBM alerted that SGR J1935+2154 was entering a new active bursting phase, Insight--HXMT began observing SGR J1935+2154 and detected 11 bursts within a total of 17 hours. The brightest burst was detected on 2020 April 28 at a geocentric arrival time of UTC 14:34:24.0114, less than a second before FRB 200428 was detected by CHIME and STARE2. All three X-ray telescopes aboard Insight--HXMT detected the burst with a combined fluence of $63.68 \pm 6.62 \times 10^{-8}$ erg cm$^{-2}$ s$^{-1}$ in the energy range of 1--250 keV, and a duration of 0.53 s \citep{li_hxmt_2021}.

Data catalogs of an Insight--HXMT extended 33-day observation of SGR J1935+2154 beginning on 2020 April 28 reports the detection of two XRBs during simultaneous monitoring of the source by FAST. The detections occurred on 2020 May 6 UTC 22:48:21.550 and 2020 May 12 UTC 21:47:43.340 \citep{2022ApJS..260...24C}. FAST detected no radio emission concurrent with the detection times.

Since SGR J1935+2154 re-entered an active bursting phase in 2022 October,  Insight-HXMT has been observing SGR J1935+2154 since 13 UTC 04:51:43. On 2022 October 21, Insight--HXMT detected an XRB associated with a radio burst detected by the Yunnan 40-meter telescope. When corrected for frequency-dependent dispersion delay, the X-ray arrival time of UTC 10:01:45 is temporally coincident with the radio arrival time \citep{2022ATel15708....1L}.

\subsubsection{NICER}

NICER was launched on 2017 June 3 to be fitted as an external attached payload on the International Space Station (ISS). NICER's sole science instrument, the X-ray Timing Instrument (XTI), is a non-imaging soft X-ray (0.2--12 keV) timing and spectroscopy instrument consisting of 56 optical modules that altogether provide an effective area of nearly 2,500 cm$^2$ at 1.5 keV \citep{2016SPIE.9905E..1HG}.

NICER began observing SGR J1935+2154 on 2020 April 28 UTC 00:40:58, 6 hrs after the initial Swift/BAT and Fermi/GBM triggers that signaled the start of an active bursting phase, and just under 14 hrs prior to FRB 200428. During the first 1120 s, NICER detected over 217 bursts during a highly active burst storm before the detection rate dropped by a factor of over 25 and remained comparatively low thereafter. During the first $\sim$600 seconds of observations while FAST was simultaneously monitoring SGR J1935+2154, NICER detected 137 bursts, 104 of which were unique from the Fermi/GBM detections \citep{younes_nicer_2020}.

\subsubsection{GECAM}
The Gravitational Wave High-energy Electromagnetic Counterpart All-sky Monitor (GECAM) is a pair of satellites developed by the Chinese Academy of Science launched on December 10, 2020. The two satellites are located on opposite sides of the Earth in order to get a simultaneous view of the entire sky. They have identical on-board instrumentation consisting of gamma-ray and charged particle detectors. GECAM's primary mission goal is to find and monitor gamma radiation from gravitational wave event sources, but it will also monitor a variety of other high-energy transients. 

On 2020 October 14 UTC 19:21:39.100, GECAM detected an XRB from SGR J1935+2154 at a de-dispersed arrival time consistent with a radio burst detected by CHIME. The in-flight and ground localization given by both instruments is consistent with SGR J1935+2154 within error \citep{2022ATel15682....1W}.

\subsection{Radio observations}
\label{sec:observations_radio}
\begin{table*}
    \caption{Radio detections from SGR J1935+2154. The first three entries are independent detections of FRB 200428. For completeness, the two sub-bursts of the CHIME detection are listed separately.}
    \label{tab:radio_observations}
    \centering
    \begin{threeparttable}
    \begin{tabular}{l|l|l|l|l|l|l}
    \hline
    \multicolumn{1}{l}{Telescope} & \multicolumn{1}{l}{Date} & \multicolumn{1}{l}{Time (UTC)} & \multicolumn{1}{c}{Fluence} & \multicolumn{1}{c}{FWHM (ms)} & \multicolumn{1}{c}{DM (pc cm$^{-3}$)} & \multicolumn{1}{c}{$\nu_0^{\mathsection}$} \\
        \hline
        CHIME$^{\dagger}$ & 2020 April 28 & 14:34:24.42650 & 480 kJy ms\tnote{$\ddagger$} & 0.585 $\pm$ 0.014 ms & 332.7 $\pm$ 0.0009 & 600 MHz\\
        CHIME$^{\dagger}$ & 2020 April 28 & 14:34:24.45547 & 220 kJy ms\tnote{$\ddagger$} & 0.335 $\pm$ 0.007 ms & 332.7 $\pm$ 0.0009 & 600 MHz\\
        STARE2$^{\dagger}$ & 2020 April 28 & 14:34:24.45548 & 1.5 $\pm$ 0.3 MJy ms & 0.61 $\pm$ 0.09 ms & 332.7 $\pm$ 0.008 & 1378 MHz\\
        \hline
        FAST & 2020 April 30 & 22:20:00 & 60 mJy ms\tnote{*} & 0.93 ms\tnote{*} & 332.9\tnote{*} & 1250 MHz\\
        Westerbork & 2020 May 24 & 22:19:19.67464 & 112 $\pm$ 22 Jy ms & 0.427 $\pm$ 0.033 ms & 332.9 $\pm$ 0.21 & 1324 MHz\\
        Westerbork & 2020 May 24 & 22:19:21.07058 & 24 $\pm$ 5 Jy ms & 0.219 $\pm$ 0.027 ms & 332.9 $\pm$ 0.21 & 1324 MHz\\
        MNC & 2020 May 30 & 00:31:03 & 457 mJy ms\tnote{*} & 114.174 ms\tnote{*} & 316.0 $\pm$ 17.5 & 408 MHz\\
        BSA/LPI & 2020 September 2 & 18:14:59 & 47.6 Jy ms\tnote{*} & 340 ms\tnote{*} & 320.0 $\pm$ 10 & 111 MHz\\
        CHIME & 2020 October 8 & 02:23:41.976 & 900 $\pm$ 160 Jy ms & 0.26 $\pm$ 0.01 ms & 332.7 $\pm$ 0.002 & 600 MHz\\
        CHIME & 2022 October 14 & 19:21:47 & 20 kJy ms$^*$ & 13 ms$^*$ & 332.8 $\pm$ 0.4 & 600 MHz\\
        Yunnan & 2022 October 21 & 10:01:45.84215 & 15 Jy ms & 0.75 ms$^*$ & 313$^*$ & 2245 MHz\\ 
        \hline
    \end{tabular}
    \begin{tablenotes}
        \item[$\mathsection$] Central observing frequency.
        \item[$\dagger$] These radio detections are collectively defined as FRB 200428.
        \item[$\ddagger$] Detection far outside the primary-beam main lobe introduces uncertainty of a factor of two \citep{chimefrb_collaboration_frb_2020}.
        \item[*] Uncertainties are not reported for these quantities.
    \end{tablenotes}
    \end{threeparttable}
\end{table*}
In the months following FRB 200428, many radio telescopes monitored SGR J1935+2154 but most observation campaigns led to null results. The few successful radio detections from SGR J1935+2154 in 2020 all appeared much weaker (and most slower) than FRB 200428. 

On 2022 October 10, Swift/BAT and INTEGRAL detected a number of XRBs from SGR J1935+2154 \citep{2022ATel15667....1P,2022GCN.32698....1M}, indicating that the magnetar was entering another active bursting phase. In the following weeks, two radio bursts were detected \citep{2022ATel15681....1D, 2022ATel15707....1H}, both with arrival times coincident with the detection of XRBs (\citep{2022ATel15682....1W, 2022ATel15708....1L}).

This section presents the detections of radio bursts from SGR J1935+2154, and their properties are listed in Table~\ref{tab:radio_observations}.

\subsubsection{STARE2}

The STARE2 is a network of three 1281--1468 MHz radio telescopes in the southwestern United States with a field-of-view of 3.6 steradians and a sensitivity threshold of 1 ms above $\sim$300 kJy \citep{bochenek_stare2_2020}.

All three STARE2 detectors were triggered at a geocentric arrival time referenced to infinite frequency of 2020 April 28 UTC 14:34:24.45548. The event, since referred to as FRB 200428, had a band-averaged fluence of 1.5 MJy ms with an effective frequency of 1,378 MHz, a FWHM temporal duration of 0.61 ms, and dispersion measure (DM) of 332.702 pc cm$^{-3}$. The STARE2 detection was temporally coincident and had a similar DM with the event detected by CHIME (discussed in Section~\ref{sec:chime_detection}), but with approximately 1,000 times higher fluence. \citep{bochenek_frb_2020}. 

\subsubsection{CHIME}
\label{sec:chime_detection}

The CHIME is a radio telescope consisting of four fixed reflecting cylinders, each with 256 equispaced antennas sensitive to 400--800 MHz radiation, altogether providing a field-of-view of at least 200 deg$^2$ \citep{2014SPIE.9145E..4VN}. Originally conceived to map baryon acoustic oscillation features in redshifted hydrogen, its large field-of-view and collecting area, wide radio bandwidth, and primary beam consisting of 1024 individual beams, make it an ideal detector of FRBs \citep{the_chimefrb_collaboration_chime_2018}.

On 2020 April 28, CHIME detected a dispersed radio burst consisting of two independent sub-bursts at geocentric arrival times referenced to infinite frequency of 14:34:24.42650(2) and 14:34:24.45547(2). The 400--800 MHz averaged fluences were 480 kJy ms and 220 kJy ms, and the temporal widths (after correcting for propagation and beam-attenuation effects) were 0.585 $\pm$ 0.014 ms and 0.335 $\pm$ 0.007 ms, respectively.  The two burst components, fitted jointly, have a dispersion measure (DM) of 332.7206(9). The values in parentheses denote uncertainties of a 68.3\% confidence interval in the last significant digit \citep{chimefrb_collaboration_frb_2020}. The second component of the CHIME burst is the lower-frequency component of the STARE2 burst since the geocentric arrival times referenced to infinite frequency are temporally consistent between the two \citep{bochenek_frb_2020}.

On 2020 October 8, CHIME detected three more radio pulses from SGR J1935+2154. The pulses were detected at UTC 02:23:41.976, 02:23:43.922 and 02:23:44.871 at the CHIME location at a frequency of 400.195 MHz. The 400--800 MHz fluences were 900 $\pm$ 160, 9.2 $\pm$ 1.6 and 6.4 $\pm$ 1.1 Jy ms. Only the first and brightest burst was fitted to obtain a DM of 332.658 $\pm$ 0.002 pc cm$^{-3}$ and an intrinsic width of 0.26 $\pm$ 0.01 ms \citep{2020ATel14080....1P}.

\subsubsection{FAST}
\label{sec:fast}

FAST is located in a radio-quiet region in southwest China and is the world's largest and most sensitive radio telescope with an illuminated aperture of 300 m. The dish can achieve sky coverage of a zenith angle up to 40$^{\circ}$ and observe across a wide frequency band of 70 MHz--3 GHz using an ultra-wideband receiver or 1.05 - 1.45 GHz using a highly sensitive 19-beam receiver \citep{nan_fast_2011}. The fluence threshold of FAST is $\approx$ 10 mJy ms for a 1 ms burst, several orders of magnitude lower than any other radio telescope in operation as of 2022 \citep{lin_no_2020}.

FAST was observing SGR J1935+2154 since 2020 April 15 and began an extended observation campaign from April 28 to May 19 with a total duration of 26 hours \citep{zhu_birth_submitted}. Unfortunately FAST was not observing SGR J1935+2154 at the time of FRB 200428. On 2020 April 30, FAST detected a highly polarized radio burst from the source at UTC 21:43:00.42 (MJD 58969.9048669008). The 19-beam receiver was mounted on the telescope for observations at a central frequency of 1.25 GHz with a bandwidth of 460 MHz. The burst fluence was 60 mJy ms with a duration of 1.966 ms and the DM was 332.9 pc cm$^{-3}$ \citep{2020ATel13699....1Z}. 
This event had far weaker fluence than any other radio signal from the source by several orders of magnitude, yet was still detected with a signal-to-noise ratio (SNR) of 1092. In the months after May, FAST stopped monitoring SGR J1935+2154 due to a lack of radio activity. In August, FAST returned twice to observe the source but did not detect any radio emission \citep{zhu_birth_submitted}. In October, FAST detected periodic radio pulses from the source, $P\approx3.248$ s, with fluences up to 40 mJy ms \citep{2020ATel14084....1Z}.

\subsubsection{Westerbork RT1}
\label{sec:wb}

Originally constructed in Netherlands in the 1960s, the Westerbork Synthesis Radio Telescope (WSRT) consists of 10 polar mounted reflector antennas each with a 25 m diameter, placed at an interval of 144 m along an east--west baseline. A 300 m long rail track at the eastern end carries two additional mobile antennas. 1.4 km to the east lies two more antennas placed on movable rails for a total of 14 identical radio antennas \citep{baars_wsrt_concept_2019}. 

Since the announcement of FRB 200428, the single 25 m dish RT1 was used alongside the 25 m and 20 m telescopes at Onsala Space Observatory in Sweden and the 32 m dish in Toruń, Poland to observe SGR J1935+2154. Between 2020 April 29 and 2020 July 27 the group of telescopes observed the source for a total on-source time of 522.7 hours. Westerbork RT1 detected two bursts at barycentric arrival times referenced to infinite frequency of UTC 22:19:19.67464 and 22:19:21.07058. The two bursts were detected at a central observing frequency of 1,324 MHz and bandwidth of 128 MHz with fluences of 112 $\pm$ 22 and 24 $\pm$ 5 and pulse widths of 427 $\pm$ 33 and 219 $\pm$ 27, respectively. The DM of the bursts were calculated to be 332.85 $\pm$ 0.21 pc cm$^{-3}$ and 332.94 $\pm$ 0.21 pc cm$^{-3}$, respectively \citep{kirsten_detection_2021}.

\subsubsection{Medicina Northern Cross}
\label{sec:mnc}

The Northern Cross at the Medicina Radio Astronomical Station (or Medicina Northern Cross; MNC) located in Bologna, Italy, is a T-shaped interferometer operating at a frequency of 408 MHz with a 16 MHz bandwitdh. The north-south arm recently underwent refurbishment for use in the search for FRBs \citep{locatelli_mnc_2020}.

Beginning on 2020 April 30, MNC monitored SGR J1935+2154 almost daily for up to 1.5 hours. On 2020 May 30, MNC detected periodic radio pulsations from the source with a SNR of 6.8. The flux density was inferred to be ~4 mJy and the pulse width was measured to be 114.174 ms, implying a fluence of 457 mJy ms. The DM was calculated to be 316.011 $\pm$ 17.526 pc cm$^{-3}$ \citep{2020ATel13783....1B}.

\subsubsection{BSA/LPI}

The Big Scanning Array of Lebedev Physical Institute (BSA/LPI), located in Puschino, Russia, is a phased antenna array operating at a central observing frequency of 111 MHz with a bandwith of 2.5 MHz \citep{tyulbashev_bsalpi_2016}. The team carried out a search of radio pulses from SGR J1935+2154 on observational data spanning 2020 January 1 to 2020 November 16, totaling approximately 26 hours of observations. They found a single pulse registered on 2020 September 2 at UTC 18:14:59 with a flux density of 140 mJy and pulse width of 2.2 s, corresponding to a fluence of $\sim$308 Jy ms. The burst was detected with a SNR of 6.6 and the DM was calculated to be 320 $\pm$ 10 pc cm$^{-3}$. The pulse arrival time indicated that the signal landed on the side lobe of the BSA LPI beam, so the flux density represents a lower limit \citep{2020ATel14186....1A}.

\subsubsection{Yunnan Astronomical Observatory}
Triggered by renewed bursting activity from SGR J1935+2154 in 2022 October, The 40-meter radio telescope at the Yunnan Astronomical Observatory in southwest China detected a radio burst on 2022 October 21 UTC 10:01:45.84215. The signal was detected at a central frequency of 2.245 GHz and frequency bandwidth of 110 MHz, with a detection significance of 20$\sigma$. The estimated flux and fluence of the burst are 20 Jy and 15 Jy ms, respectively. The best estimation for the DM is 313 pc cm$^{-3}$, which is significantly lower than previously reported detections, which indicates a dynamically evolving environment \citep{2022ATel15707....1H}.

\section{Results and Discussion}
\label{sec:results}

The radio detections from SGR J1935+2154 (Table~\ref{tab:radio_observations}) span a period of approximately 5 months during an active bursting phase of the Galactic magnetar wherein several radio pulses have been detected. The fast radio burst on 2020 April 28 (FRB 200428) was detected by both CHIME and STARE2 with fluence values many orders of magnitude greater than any radio signal ever detected. CHIME detected two components separated by $29$ ms, where the second component is temporally coincident with the STARE2 detection. These three independent signals are collectively identified as FRB 200428. For completeness they are included in Table~\ref{tab:radio_observations}.

Given two independent detections of FRB 200428 in different wavelength bands, and assuming the burst is of Gaussian spectral shape, one obtains a characteristic burst spectral width of FRB 200428 referenced to its central frequency, $\delta\nu_{\text{FRB}}/\nu_{\text{0,FRB}}=0.278$ (Appendix~\ref{appx:spectral_width}).

FRB 200428 was detected in association with an XRB. We therefore initially hypothesize that all SRBs are similarly associated with XRBs.

However, some radio bursts were detected with an associated XRB while most were not. During the 2020 active bursting phase of SGR J1935+2154, no radio pulse besides FRB 200428 was detected with an associated XRB. Furthermore, they were detected with much lower fluences and longer durations than FRB 200428. Meanwhile, in 2020, FAST detected no pulsed radio emission from the source concurrent with 140 XRBs before and after FRB 200428. During the 2022 active bursting phase, two radio pulses were detected with associated XRBs.
We use these observational facts to constrain FRB beaming under our first hypothesis.

\begin{figure}
    \centering
    \includegraphics[scale=0.5]{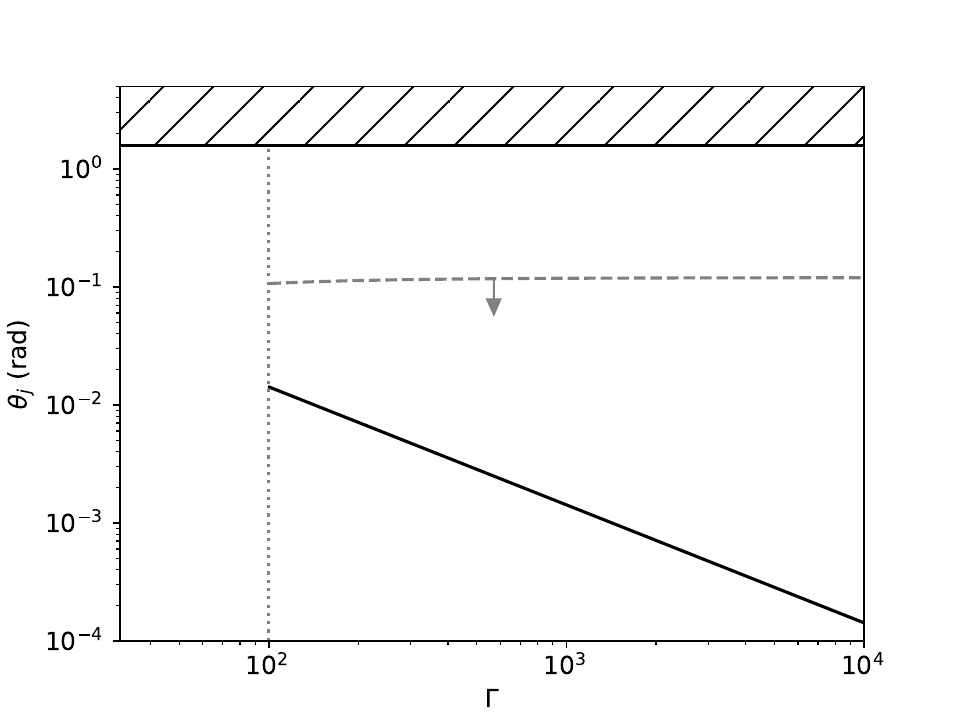}
    \caption{Observational beaming constraints of FRBs from the Galactic magnetar SGR J1935+2154. The main constraint (\textit{solid, black}) is obtained from the detection of two SRBs and one FRB from the source ($\mathcal{R}_{\Delta\Omega}=2$). An upper limit constraint (\textit{dashed, black}) is obtained by non-detections of radio emission by FAST concurrent with 140 XRBs, where every XRB is assumed to be intrinsically associated with an SRB ($f_{\text{FRB}}=1$). Note that the main constraint is consistent with the upper limit. If not every XRB is assumed to be intrinsically associated with an SRB ($f_{\text{FRB}}<1$), the upper limit is raised further and remains consistent with the main constraint.}
    \label{fig:frb_xrb}
\end{figure}

\subsection{XRB-associated SRBs}
\label{sec:srb_xrb_results}

A total of three radio pulses have been detected in association with XRBs: FRB 200428 in 2020 and the two radio pulses in 2022. The two radio pulses in 2022 had much weaker fluence and longer duration than FRB 200428, suggesting they might be SRBs. We use the FRB--SRB closure relation of equation~\ref{eq:frb_srb_closure_gaussian} with a reference FRB identical to FRB 200428 (hereafter "reference FRB-1"). For the CHIME radio detection on 2022 October 14 there is no solution, meaning that it cannot be an SRB of a reference FRB-1 with a Gaussian-like spectrum. We instead use the power-law closure relation derived in \cite{zhang_slow_2021} to find that the CHIME detection on 2022 October 14 may indeed be an SRB of a reference FRB-1 with a power-law spectrum of index $\alpha=-0.8$. For the Yunnan radio detection on 2022 October 21, we find that it may be an SRB of a reference FRB-1 with a Gaussian spectrum of narrow characteristic width, $\delta\nu/\nu = 0.16$. 

The detection of two SRBs and one FRB from SGR J1935+2154 implies that the ratio of solid angles between SRBs and FRBs is $\mathcal{R}_{\Delta\Omega}\sim2$. Equation~\ref{eq:r_sa} represents how the solid angle ratio depends on geometric and relativistic FRB beaming factors, which is used to constrain observations. The observational constraints are given in Fig.~\ref{fig:frb_xrb}. We find that FRB beaming must be geometrically narrow and that the beaming factors follow $\theta_j\Gamma\sim2$.

An additional constraint on FRB beaming is imposed by the non-detections of radio emission by FAST concurrent with 140 XRBs. We assign an upper limit for FAST detecting an FRB associated with an XRB as $P \leq 1/141$. This upper-limit constraint is consistent with the main constraint. Theoretical descriptions of FRBs derive that the emission is highly relativistic, $\Gamma\gtrsim10^2$ \citep{lyubarsky_08,murase_16,lu_kumar_16,Zhang_ICS_22}, see \cite{zhang_RMP_22} for a review. We additionally constrain the relativistic beaming factor to be $\Gamma\geq10^2$.

\subsubsection{On the possible uniqueness of SRB-associated XRBs}
\label{sec:f_frb_results}

During its 2020 observation campaigns of SGR J1935+2154, FAST detected no radio emission concurrent with 140 XRBs. The utter lack of radio detections by the highly sensitive FAST challenges the assumption that all XRBs are intrinsically associated with SRBs. Furthermore, recent studies suggest that the XRB associated with FRB 200428 was intrinsically unique from the majority \citep{2021NatAs...5..408Y,yang_bursts_2021,li_quasi-periodic_2022}. We consider the possibility that not all XRBs are intrinsically associated with SRBs by introducing a free parameter, $f_{\text{FRB}} \leq 1$, that represents the fraction of XRBs that are intrinsically associated with FRBs (SRBs). If all SGR bursts that produce XRBs also produce FRBs, $f_{\text{FRB}} = 1$, and the earlier constraints are unchanged. On the other hand, if not all XRBs are intrinsically associated with FRBs, $f_{\text{FRB}} < 1$.

In this work we consider the limiting case of $f_{\text{FRB}}=1/141$ based on observations by FAST. While we now assume that one out of 141 XRBs are intrinsically associated with FRBs, we still assume that all FRBs (SRBs) are associated with XRBs. That is, the relationship between FRBs and SRBs is unaffected by $f_{\text{FRB}}$ since both are equally subject to the added constraint, and so $\mathcal{R}_{\Delta\Omega}$ is unchanged. The FRB beaming constraint by non-detections of radio emission by FAST concurrent with XRBs (see Fig.~\ref{fig:frb_xrb}) becomes $P \leq 1/(141 \times f_{\text{FRB}}) = 1$ and the geometric constraint is loosened to the maximum value, $\theta_j=\pi/2$. This is expected since the introduction of $f_{\text{FRB}} < 1$ ascribes the observational scarcity to more than just beaming (i.e., intrinsic factors) so that constraints by beaming alone must be weaker. We note that regardless of the value of $f_{\text{FRB}}$, the upper-limit constraint for FRB beaming remains consistent with the main constraint.

\subsection{XRB-independent SRBs}
\label{sec:frb_srb_results}

\begin{table*}
    \centering
    \caption{SRBs of SGR J1935+2154 by solution of the FRB--SRB closure relation. Bolded entries denote SRBs detected in associated with an XRB.}
    \label{tab:frb_srb}
    \begin{threeparttable}
    \begin{tabular}[t]{llllll}
    \hline
    \multicolumn{1}{l}{Reference FRB} &
    \multicolumn{1}{l}{Scope} &
    \multicolumn{1}{l}{Date} &
    \multicolumn{1}{l}{Time (UTC)} & \multicolumn{1}{l}{$\delta\nu/\nu^{\dagger}$} &
    \multicolumn{1}{l}{$\alpha^{\ddagger}$}\\
    \hline
    \multirow{3}{*}{\textit{FRB 200428}}
    & FAST & 2020 April 30 & 21:43:00.42 & 0.36 & ---\\
    & BSA/LPI & 2020 September 2 & 18:14:59 & --- & -1.1\\
    & \textbf{Yunnan} & \textbf{2022 October 21} & \textbf{10:01:45.84215} & \textbf{0.16} & ---\\
    \hline
    \multirow{4}{*}{\shortstack[l]{\textit{Short-Duration} \\\textit{Cosmological FRB}}}
    & Westerbork & 2020 May 24 & 22:19:19.67464 & 0.80 & ---\\
    & Westerbork & 2020 May 24 & 22:19:21.07058 & 0.27 & ---\\
    & CHIME & 2020 October 8 & 02:23:41.976 & 0.068 & ---\\
    & \textbf{CHIME} & \textbf{2022 October 14} & \textbf{19:21:47} & --- & \textbf{-0.8}\\
    \hline
    \end{tabular}
    \begin{tablenotes}
        \item[$\dagger$]{Gaussian spectral width referenced to peak of the associated FRB.}
        \item[$\ddagger$]{Power-law index of the associated FRB.}
    \end{tablenotes}
    \end{threeparttable}
\end{table*}

In our second hypothesis we eliminate the requirement that SRBs must be associated with XRBs. We again use the FRB--SRB closure relation in equation~\ref{eq:frb_srb_closure_gaussian} to determine whether any of the subsequent radio detections are SRBs. We first consider a reference FRB identical to FRB 200428 ("reference FRB-1").

\subsubsection{SRBs of FRB 200428}
\label{sec:frb_srb_200428_results}
The MNC detection satisfies the FRB--SRB closure relation with reference FRB-1 if the FRB has a wide spectrum, $\delta\nu/\nu=15$. There is no solution found for the BSA/LPI detection, meaning that it cannot be an SRB of reference FRB-1 if the FRB spectrum is a Gaussian. \cite{zhang_slow_2021} showed that the burst detected by BSA/LPI can be an SRB of reference FRB-1 only if the FRB spectrum is a power-law of index $\alpha=-1.1$. The highly polarized burst detected by FAST on 2020 April 30 can be an SRB of reference FRB-1 if the FRB has a narrow spectral width, $\delta\nu/\nu=0.36$. The burst detected by the Yunnan Observatory can be an SRB of reference FRB-1 if the FRB has a narrow spectral width, $\delta\nu/\nu=0.16$. These SRBs are listed in Table~\ref{tab:frb_srb}.

One would expect that the FRB spectral shape is generally consistent between a population of FRBs from the same source. We note that the average spectral width of reference FRB-1 required by the observed SRBs, $\delta\nu/\nu = 0.26$, is roughly consistent with the characteristic spectral width of FRB 200428 itself, $\delta\nu/\nu = 0.278$ (Appendix~\ref{appx:spectral_width}). On the contrary, the BSA/LPI burst may be an SRB of reference FRB-1 only if the FRB has a power-law spectrum. We attribute the inconsistencies in spectral shape to intrinsic factors that are not addressed here. We note that such behavior is observed in the FRB population as a whole, with broad- and narrow-band spectra being regularly observed \citep[e.g.,][]{aggarwal_frb121102_2021,2020ATel14080....1P}, even among different bursts for the same source \citep{zhou22}.

The MNC "burst" was in fact a marginal detection of radio pulsations rather than a single bursting event \citep{2020ATel13783....1B}. Furthermore, all potential SRBs discussed in this work require their reference FRB to have either a narrow Gaussian spectrum with $\delta\nu/\nu<1$ or power-law spectrum with $\alpha \sim -1$; only the MNC burst demands legitimacy as an SRB of reference FRB-1 with a wide Gaussian spectrum. For these reasons we exclude the MNC burst from consideration as an SRB.

\subsubsection{SRBs of a short-duration cosmological FRB}
\label{sec:frb_srb_short_results}

There have been several radio detections from SGR J1935+2154 with shorter duration than FRB 200428. By definition, these radio pulses cannot be SRBs of reference FRB-1. It is instead possible that they may be SRBs of a different reference FRB. In this section we consider a short-duration cosmological reference FRB, $\mathcal{F_{\nu}}=100 \text{ MJy ms}$ and $w=0.1 \text{ ms}$ (hereafter "reference FRB-2").

The two radio detections by Westerbork on 2020 May 24 UTC 22:19:19 and 22:19:21 may both be SRBs of reference FRB-2 if the FRBs have narrow Gaussian spectra, $\delta\nu/\nu=0.80$ and $\delta\nu/\nu=0.27$, respectively. The CHIME detection on 2020 October 8 may be an SRB of reference FRB-2 if the FRB has a narrower Gaussian spectrum, $\delta\nu/\nu=0.068$. The CHIME detection on 2022 October 14 may be an SRB of reference FRB-2 if the FRB has a power-law spectrum, $\alpha=-0.8$. These SRBs are listed in Table~\ref{tab:frb_srb}.

As with reference FRB-1, we again distinguish between SRBs based on the required spectral shape of reference FRB-2. The three radio detections in 2020 all require that reference FRB-2 has a narrow Gaussian spectrum. We note without statistical justification that the mean spectral width of reference FRB-2 required by these SRBs, $\delta\nu/\nu=0.38$, is roughly consistent with the characteristic spectral width of FRB 200428 itself, $\delta\nu/\nu=0.278$ (Appendix~\ref{appx:spectral_width}), as well as the average spectral width of reference FRB-1 required by the SRBs detected by FAST and Yunnan Observatory, $\delta\nu/\nu=0.26$. The power-law SRB detected by CHIME on 2022 October 14 is thus distinguished from the three Gaussian SRBs detected in 2020, as in the case of the BSA/LPI burst and reference FRB-1.

We note that the three bursts discussed in Section~\ref{sec:frb_srb_200428_results} can also be SRBs of reference FRB-2 solely due to their longer duration. Then, the spectra of reference FRB-2 must be very wide, $\delta\nu/\nu=3.5$ and $\delta\nu/\nu=2.3$ for the FAST and Yunnan Observatories detections, respectively. We deem it unnecessary to consider reference FRB-2 instead of reference FRB-1 in these cases for two reasons. First, one should expect FRBs from a Galactic magnetar to be similar to FRB 200428 (i.e., not as bright as a cosmological FRB). Second, the spectral widths obtained by using reference FRB-1 are more consistent with the majority of FRBs and SRBs discussed.

\subsubsection{FRB beaming constraints}
\label{sec:frb_srb_constraint_results}

\begin{figure}
    \centering
    \includegraphics[scale=0.5]{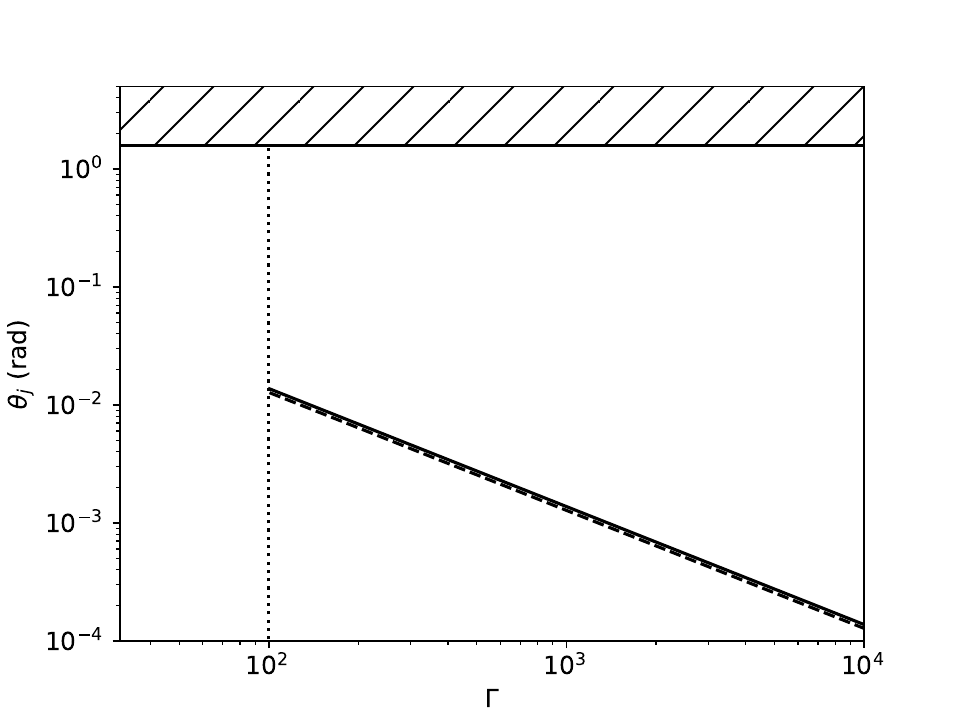}
    \caption{
    Observational beaming constraints of FRBs from the Galactic magnetar SGR J1935+2154. For reference FRB-1 with a Gaussian spectrum ($\delta\nu/\nu=0.26$), there are two associated SRBs, $\mathcal{R}_{\Delta\Omega} = 2$ (\textit{solid}). For reference FRB-2 with a Gaussian spectrum ($\delta\nu/\nu=0.38$), there are three associated SRBs, $\mathcal{R}_{\Delta\Omega} = 3$ (\textit{dashed}). The hatched region represents the geometrically forbidden and unphysical region of FRB beaming, $\theta_j>\pi/2$. The dotted vertical line indicates the theoretical relativistic lower limit of FRBs, $\Gamma \gtrsim 10^2$.}
    \label{fig:frb_srb_observation}
\end{figure}

Under our second hypothesis we obtain a total of seven SRBs, five of which were not detected with an associated XRB. All SRBs are listed in table~\ref{tab:frb_srb} along with the spectral parameters of their reference FRBs obtained from the FRB--SRB closure relations. Three SRBs are found to be related to a reference FRB identical to FRB 200428, two of which require it has a narrow Gaussian spectrum. Considering only the SRBs of FRBs with Gaussian spectra, $\mathcal{R}_{\Delta\Omega} = 2$. The solid line in Fig.~\ref{fig:frb_srb_observation} shows the geometric and relativistic beaming constraints for reference FRB-1 given a Gaussian spectral width equal to the average required by the two related SRBs, $\delta\nu/\nu=0.26$.

Meanwhile, four SRBs are found to be related to a short-duration cosmological reference FRB, three of which require that it has a narrow Gaussian spectrum. Considering only the SRBs of FRBs with Gaussian spectra, $\mathcal{R}_{\Delta\Omega} = 3$. The dashed line in Fig.~\ref{fig:frb_srb_observation} shows the geometric and relativistic beaming constraints for reference FRB-2 given a Gaussian spectral width equal to the average required by the three related SRBs, $\delta\nu/\nu=0.38$.

The derived beaming constraints for the two reference FRBs are nearly identical. This coincidence is largely due to $\mathcal{R}_{\Delta\Omega}$ being of order unity for either reference FRB, but is more precisely due to the particular properties of each reference FRB (i.e., fluence, pulse duration, Gaussian spectral width). We find that the geometric and relativistic beaming factors for both reference FRBs follow $\theta_j\Gamma\sim2$. As more Galactic SRBs and FRBs are detected, $\mathcal{R}_{\Delta\Omega}$ will fluctuate and converge on the physically accurate solid angle ratio between SRBs and FRBs.

\section{Conclusions}
\label{sec:conclusions}

In this work we extended theoretical models to constrain geometric and relativistic beaming factors of FRBs with Gaussian spectra and developed computer code to numerically solve the equations. Our results indicate that FRBs with narrow Gaussian spectra are more difficult to geometrically and relativistically constrain. This is due to the fact that narrow spectra are more sensitive to relativistic effects versus wide FRB spectra. We provided observational constraints under two separate hypotheses.

In the first we assumed that SRBs are associated with XRBs like the case of FRB 200428 and find a total of two SRBs.
We constrain FRB beaming to be geometrically narrow and highly relativistic, with combined beaming factors that follow $\theta_j\Gamma\sim2$. 

The observational scarcity of FRBs associted with XRBs, along with findings from recent studies, suggest that the XRB associated with FRB 200428 was intrinsically unique. We investigated how the beaming interpretation is affected by this uncertainty by introducing the free parameter, $f_{\text{FRB}}$ to represent the fraction of XRBs that are intrinsically associated with FRBs. We note that $f_{\text{FRB}}$ only affects the constraint obtained by non-detections of radio emission by FAST concurrent with XRBs. As radio-burst-associated XRBs are considered increasingly rare, $f_{\text{FRB}} < 1$, the geometric constraint on FRB beaming is increasingly loosened. At a reference limit of $f_{\text{FRB}}=1/141$, the geometric constraint is loosened to the maximum value, $\theta_j=\pi/2$. We confirmed that for any value of $f_{\text{FRB}}$, the upper-limit constraint on FRB beaming remains consistent with the main constraint.

In our second hypothesis we forgo the stringent requirement that all SRBs are associated with XRBs. We find an additional five potential SRBs from SGR J1935+2154 for a grand total of seven SRBs from the Galactic magnetar. Until a considerable number of additional detections are successful, $\mathcal{R}_{\Delta\Omega}\sim2$ and FRB beaming factors will follow $\theta_j\Gamma \sim 2$.

Regardless of whether FRBs and SRBs are intrinsically associated with XRBs, our results under both hypotheses show that FRB beaming is geometrically narrow and highly relativistic. This fact is consistent with theoretical models for FRBs that support a magnetospheric origin of emission \citep{Kumar17, Yang_18, metzger_19, Wadiasingh_2019, wang_magnetospheric_2020, kumar_20, lu_unified_2020, yang_pair_20, Wadiasingh_2020, Qu21, Yang_21, Wang_22, Zhang_ICS_22}. Further FRB, SRB and XRB detections from SGR J1935+2154 will continually improve constraints on FRB beaming until statistically precise values can be established.

\section*{Acknowledgements}
This work is supported by the Nevada Center for Astrophysics.

\section*{Data Availability}

The Python code developed to perform calculations as part of this work is available upon request.



\bibliographystyle{mnras}
\bibliography{frb_srb}



\appendix
\section{X-ray detections from SGR J1935+2154}
\label{appx:xray_observations}
Table~\ref{tab:xray_observations_radio} lists the XRBs from SGR J1935+2154 detected in association with radio emission. Table~\ref{tab:xray_observations_fast} lists the XRBs detected during FAST observing windows of the same source.

\begin{table}
    \def\arraystretch{1.2}
    \setlength{\tabcolsep}{5.5pt}
    \centering
    \caption{X-ray bursts from SGR J1935+2154 detected in association with radio emission.}
    \label{tab:xray_observations_radio}
    \begin{threeparttable}
    \begin{tabular}[t]{llllll}
    \hline
    \multicolumn{1}{l}{\#} &
    \multicolumn{1}{l}{Telescope} &
    \multicolumn{1}{l}{Date} &
    \multicolumn{1}{l}{Time (UTC)} & \multicolumn{1}{l}{$T_{90}$ (s)} &
    \multicolumn{1}{l}{Flux$^{\dagger}$}\\
    \hline
    1a & INTEGRAL$^1$ & 2020-4-28 & 14:34:24.357 & 0.75 & $10.2^{+0.5}_{-0.5}$ \\
    1b & HXMT$^2$ & 2020-4-28 & 14:34:24.011 & 1.2 & $7.14^{+0.41}_{-0.38}$ \\
    1c & KW$^3$ & 2020-4-28 & 14:34:24.447 & 0.464 & $9.7^{+1.1}_{-1.1}$ \\
    1d & AGILE$^4$ & 2020-4-28 & 14:34:24.400 & 0.5 & $\sim10$ \\
    2 & GECAM$^5$ & 2022-10-14 & 19:21:39.100 & 0.25 & ---$^*$ \\
    3 & HXMT & 2022-10-21 & 10:01:45 & ---$^*$ & ---$^*$ \\
    \hline
    \end{tabular}
    \begin{tablenotes}
        \item[$\dagger$]{(10$^{-7}$ erg cm$^{-2}$ s$^{-1}$).}
        \item[*]{Data not yet available due to recency of observations at the time of writing.}
        \item[1]{INTEGRAL detections are provided in the energy range of 20--200 keV \citep{mereghetti_integral_2020}.}
        \item[2]{HXMT detections are provided in the energy range of 1--250 keV \citep{li_hxmt_2021}.}
        \item[3]{Konus-Wind detections are provided in the energy range of 20--500 keV \citep{ridnaia_kw_2021}.}
        \item[4]{AGILE detections are provided in the energy range of 18--60 keV \citep{tavani_agile_2021}.}
        \item[5]{GECAM detections are provided in the energy range of 20--100 keV \citep{2022ATel15682....1W}.}
    \end{tablenotes}
    \end{threeparttable}
\end{table}

\begin{table*}
    \def\arraystretch{1.2}
    \setlength{\tabcolsep}{5.5pt}
    \centering
    \caption{X-ray bursts from SGR J1935+2154 concurrent with FAST observing windows.}
    \label{tab:xray_observations_fast}
    \begin{tabular}[t]{llllll}
    \hline
    \multicolumn{1}{l}{\#} &
    \multicolumn{1}{l}{Scope} &
    \multicolumn{1}{l}{Date} &
    \multicolumn{1}{l}{Time (UTC)} & \multicolumn{1}{l}{$T_{90}$ (s)} &
    \multicolumn{1}{l}{Flux$^{\dagger}$}\\
    \hline
        1 & Fermi$^1$ & 2020-04-28 & 00:19:44.192 & 0.080 & $142.75^{+25.00}_{-21.50}$\\
        2 & Fermi & 2020-04-28 & 00:23:04.728 & 0.021 & $143.33^{+31.43}_{-29.05}$\\
        3 & Fermi & 2020-04-28 & 00:24:30.296 & 0.122 & $2858.93^{+56.48}_{-59.10}$\\
        4 & Fermi & 2020-04-28 & 00:25:43.945 & 0.076 & $16.97^{+10.66}_{-9.21}$\\
        5 & Fermi & 2020-04-28 & 00:37:36.153 & 0.095 & $52.63^{+12.00}_{-11.16}$\\
        6 & Fermi & 2020-04-28 & 00:39:39.513 & 0.194 & $107.94^{+13.45}_{-10.98}$\\
        7 & Fermi & 2020-04-28 & 00:40:33.072 & 0.190 & $188.84^{+15.21}_{-14.89}$\\
        8 & NICER$^2$ & 2020-04-28 & 00:41:21.260 & 0.194 & $0.32^{+0.18}_{-0.12}$\\
        9 & NICER & 2020-04-28 & 00:41:23.653 & 0.088 & ---$^*$\\
        10 & Fermi & 2020-04-28 & 00:41:32.136 & 0.222 & $623.83^{+20.50}_{-20.18}$\\
        11 & NICER & 2020-04-28 & 00:41:49.321 & 0.216 & $0.40^{+0.10}_{-0.08}$\\
        12 & NICER & 2020-04-28 & 00:41:53.367 & 0.067 & ---$^*$\\
        13 & NICER & 2020-04-28 & 00:41:56.329 & 1.453 & $0.15^{+0.02}_{-0.02}$\\
        14 & NICER & 2020-04-28 & 00:42:00.253 & 0.225 & $0.16^{+0.09}_{-0.06}$\\
        15 & NICER & 2020-04-28 & 00:42:01.643 & 0.538 & $2.45^{+0.12}_{-0.11}$\\
        16 & NICER & 2020-04-28 & 00:42:14.484 & 0.462 & $0.29^{+0.04}_{-0.04}$\\
        17 & NICER & 2020-04-28 & 00:42:26.739 & 0.470 & $0.10^{+0.03}_{-0.02}$\\
        18 & NICER & 2020-04-28 & 00:42:43.711 & 0.188 & $0.65^{+0.15}_{-0.12}$\\
        19 & NICER & 2020-04-28 & 00:42:48.218 & 0.249 & $0.10^{+0.03}_{-0.02}$\\
        20 & NICER & 2020-04-28 & 00:42:52.058 & 0.985 & $2.09^{+0.10}_{-0.09}$\\
        21 & NICER & 2020-04-28 & 00:42:54.467 & 0.274 & $6.31^{+0.30}_{-0.28}$\\
        22 & NICER & 2020-04-28 & 00:43:01.939 & 1.521 & $0.12^{+0.03}_{-0.02}$\\
        23 & NICER & 2020-04-28 & 00:43:10.976 & 1.923 & $0.19^{+0.02}_{-0.02}$\\
        24 & NICER & 2020-04-28 & 00:43:16.911 & 0.331 & ---$^*$\\
        25 & NICER & 2020-04-28 & 00:43:22.492 & 1.401 & $0.36^{+0.04}_{-0.02}$\\
        26 & Fermi & 2020-04-28 & 00:43:25.169 & 0.174 & $365.00^{+22.82}_{-20.63}$\\
        27 & NICER & 2020-04-28 & 00:43:33.334 & 0.809 & $0.07^{+0.02}_{-0.01}$\\
        28 & NICER & 2020-04-28 & 00:43:37.945 & 2.015 & $0.20^{+0.02}_{-0.02}$\\
        29 & NICER & 2020-04-28 & 00:43:40.490 & 0.113 & $0.20^{+0.12}_{-0.07}$\\
        30 & NICER & 2020-04-28 & 00:43:42.183 & 0.146 & ---$^*$\\
        31 & NICER & 2020-04-28 & 00:43:45.240 & 0.809 & $0.21^{+0.04}_{-0.03}$\\
        32 & NICER & 2020-04-28 & 00:44:00.180 & 1.366 & $0.17^{+0.02}_{-0.02}$\\
        33 & NICER & 2020-04-28 & 00:44:05.104 & 0.319 & $0.20^{+0.05}_{-0.04}$\\
        34 & Fermi & 2020-04-28 & 00:44:08.202 & 0.154 & $4698.12^{+81.10}_{-76.69}$\\
        35 & Fermi & 2020-04-28 & 00:44:09.302 & 0.112 & $455.98^{+26.07}_{-24.11}$\\
        36 & NICER & 2020-04-28 & 00:44:19.570 & 1.433 & $0.19^{+0.02}_{-0.02}$\\
        37 & NICER & 2020-04-28 & 00:44:26.236 & 0.955 & $0.28^{+0.03}_{-0.03}$\\
        38 & NICER & 2020-04-28 & 00:44:32.458 & 1.174 & $0.23^{+0.03}_{-0.03}$\\
        39 & NICER & 2020-04-28 & 00:44:40.056 & 0.417 & $0.16^{+0.04}_{-0.03}$\\
        40 & NICER & 2020-04-28 & 00:44:45.286 & 0.398 & $0.34^{+0.05}_{-0.04}$\\
        41 & NICER & 2020-04-28 & 00:44:48.974 & 1.041 & $0.38^{+0.04}_{-0.03}$\\
        42 & NICER & 2020-04-28 & 00:44:49.855 & 0.123 & $1.41^{+0.25}_{-0.21}$\\
        43 & NICER & 2020-04-28 & 00:44:51.209 & 0.954 & $0.10^{+0.03}_{-0.02}$\\
        44 & NICER & 2020-04-28 & 00:44:56.358 & 0.998 & $0.13^{+0.03}_{-0.02}$\\
        45 & NICER & 2020-04-28 & 00:44:59.898 & 0.300 & $2.51^{+0.18}_{-0.17}$\\
        46 & NICER & 2020-04-28 & 00:45:05.783 & 0.350 & $0.40^{+0.08}_{-0.06}$\\
        47 & NICER & 2020-04-28 & 00:45:11.175 & 0.854 & $0.85^{+0.06}_{-0.06}$\\
        48 & NICER & 2020-04-28 & 00:45:12.122 & 0.365 & $0.38^{+0.07}_{-0.06}$\\
        49 & NICER & 2020-04-28 & 00:45:20.845 & 0.145 & ---$^*$\\
        \hline
    \end{tabular}
    \begin{tabular}[t]{llllll}
    \hline
    \multicolumn{1}{l}{\#} &
    \multicolumn{1}{l}{Scope} &
    \multicolumn{1}{l}{Date} &
    \multicolumn{1}{l}{Time (UTC)} & \multicolumn{1}{l}{$T_{90}$ (s)} &
    \multicolumn{1}{l}{Flux$^{\dagger}$}\\
    \hline
        50 & NICER & 2020-04-28 & 00:45:21.543 & 1.259 & $0.25^{+0.02}_{-0.02}$\\
        51 & NICER & 2020-04-28 & 00:45:24.151 & 1.131 & $0.31^{+0.03}_{-0.03}$\\
        52 & NICER & 2020-04-28 & 00:45:28.849 & 0.915 & $0.21^{+0.03}_{-0.03}$\\
        53 & Fermi & 2020-04-28 & 00:45:31.097 & 0.030 & $213.67^{+32.67}_{-30.33}$\\
        54 & NICER & 2020-04-28 & 00:45:33.729 & 0.921 & $0.13^{+0.03}_{-0.03}$\\
        55 & NICER & 2020-04-28 & 00:45:39.254 & 0.655 & $1.55^{+0.11}_{-0.10}$\\
        56 & NICER & 2020-04-28 & 00:45:42.233 & 1.484 & $0.08^{+0.01}_{-0.01}$\\
        57 & NICER & 2020-04-28 & 00:45:43.978 & 0.259 & $0.58^{+0.10}_{-0.07}$\\
        58 & NICER & 2020-04-28 & 00:45:46.842 & 2.015 & $0.50^{+0.04}_{-0.03}$\\
        59 & NICER & 2020-04-28 & 00:45:48.460 & 0.777 & $0.13^{+0.03}_{-0.02}$\\
        60 & NICER & 2020-04-28 & 00:45:49.738 & 0.513 & $0.49^{+0.06}_{-0.05}$\\
        61 & NICER & 2020-04-28 & 00:45:51.765 & 0.289 & $0.20^{+0.04}_{-0.03}$\\
        62 & NICER & 2020-04-28 & 00:45:56.514 & 0.962 & $0.27^{+0.04}_{-0.03}$\\
        63 & NICER & 2020-04-28 & 00:45:57.347 & 0.088 & $3.24^{+0.57}_{-0.48}$\\
        64 & NICER & 2020-04-28 & 00:45:58.064 & 0.544 & $0.25^{+0.04}_{-0.03}$\\
        65 & Fermi & 2020-04-28 & 00:46:00.009 & 0.208 & $113.17^{+13.80}_{-11.20}$\\
        66 & Fermi & 2020-04-28 & 00:46:00.609 & 0.126 & $232.54^{+21.27}_{-18.25}$\\
        67 & Fermi & 2020-04-28 & 00:46:06.408 & 0.019 & $228.42^{+44.74}_{-39.47}$\\
        68 & NICER & 2020-04-28 & 00:46:08.257 & 0.602 & $0.15^{+0.03}_{-0.02}$\\
        69 & NICER & 2020-04-28 & 00:46:12.379 & 1.277 & $0.08^{+0.01}_{-0.01}$\\
        70 & NICER & 2020-04-28 & 00:46:14.233 & 0.840 & $0.24^{+0.03}_{-0.03}$\\
        71 & NICER & 2020-04-28 & 00:46:18.015 & 2.852 & $1.82^{+0.04}_{-0.04}$\\
        72 & Fermi & 2020-04-28 & 00:46:20.176 & 0.166 & $2881.93^{+52.23}_{-55.42}$\\
        73 & Fermi & 2020-04-28 & 00:46:23.504 & 0.742 & $38.49^{+4.95}_{-5.07}$\\
        74 & NICER & 2020-04-28 & 00:46:27.415 & 1.129 & $0.25^{+0.02}_{-0.02}$\\
        75 & NICER & 2020-04-28 & 00:46:29.769 & 1.263 & $1.38^{+0.07}_{-0.06}$\\
        76 & NICER & 2020-04-28 & 00:46:33.658 & 3.700 & $1.32^{+0.03}_{-0.03}$\\
        77 & NICER & 2020-04-28 & 00:46:40.541 & 1.906 & $0.42^{+0.03}_{-0.03}$\\
        78 & Fermi & 2020-04-28 & 00:46:43.208 & 0.128 & $265.86^{+20.00}_{-19.38}$\\
        79 & NICER & 2020-04-28 & 00:46:46.829 & 1.450 & $0.20^{+0.02}_{-0.02}$\\
        80 & NICER & 2020-04-28 & 00:46:48.887 & 0.697 & $0.29^{+0.05}_{-0.04}$\\
        81 & NICER & 2020-04-28 & 00:46:50.557 & 0.923 & $0.45^{+0.04}_{-0.04}$\\
        82 & NICER & 2020-04-28 & 00:46:56.705 & 2.204 & $0.87^{+0.04}_{-0.04}$\\
        83 & NICER & 2020-04-28 & 00:46:59.743 & 0.904 & $1.55^{+0.07}_{-0.07}$\\
        84 & NICER & 2020-04-28 & 00:47:02.017 & 1.086 & $0.48^{+0.05}_{-0.04}$\\
        85 & NICER & 2020-04-28 & 00:47:04.505 & 1.721 & $1.05^{+0.05}_{-0.05}$\\
        86 & NICER & 2020-04-28 & 00:47:09.756 & 2.044 & $0.17^{+0.02}_{-0.02}$\\
        87 & NICER & 2020-04-28 & 00:47:14.611 & 4.472 & $0.52^{+0.02}_{-0.02}$\\
        88 & NICER & 2020-04-28 & 00:47:18.948 & 0.910 & $3.16^{+0.15}_{-0.14}$\\
        89 & Fermi & 2020-04-28 & 00:47:24.961 & 0.152 & $40.79^{+7.89}_{-6.58}$\\
        90 & NICER & 2020-04-28 & 00:47:27.151 & 0.409 & $0.25^{+0.05}_{-0.04}$\\
        91 & NICER & 2020-04-28 & 00:47:30.390 & 0.789 & $0.10^{+0.02}_{-0.02}$\\
        92 & NICER & 2020-04-28 & 00:47:32.145 & 0.579 & $1.91^{+0.14}_{-0.13}$\\
        93 & NICER & 2020-04-28 & 00:47:35.689 & 0.809 & $6.17^{+0.14}_{-0.14}$\\
        94 & NICER & 2020-04-28 & 00:47:39.312 & 0.861 & $0.26^{+0.03}_{-0.03}$\\
        95 & NICER & 2020-04-28 & 00:47:41.968 & 0.324 & $0.45^{+0.09}_{-0.07}$\\
        96 & NICER & 2020-04-28 & 00:47:43.459 & 0.157 & ---$^*$\\
        97 & NICER & 2020-04-28 & 00:47:44.331 & 1.941 & $0.32^{+0.02}_{-0.02}$\\
        98 & NICER & 2020-04-28 & 00:47:47.056 & 0.826 & $0.15^{+0.03}_{-0.03}$\\
        \hline
    \end{tabular}
\end{table*}

\begin{table}
\def\arraystretch{1.2}
    \centering
    \caption*{\textbf{Continued.}}
    \begin{threeparttable}
    \begin{tabular}[t]{llllll}
    \hline
    \multicolumn{1}{l}{\#} &
    \multicolumn{1}{l}{Scope} &
    \multicolumn{1}{l}{Date} &
    \multicolumn{1}{l}{Time (UTC)} & \multicolumn{1}{l}{$T_{90}$ (s)} &
    \multicolumn{1}{l}{Flux$^{\dagger}$}\\
    \hline
        99 & NICER & 2020-04-28 & 00:47:52.328 & 0.719 & $3.31^{+0.16}_{-0.15}$\\
        100 & NICER & 2020-04-28 & 00:47:56.536 & 0.136 & $0.32^{+0.18}_{-0.07}$\\
        101 & Fermi & 2020-04-28 & 00:47:57.528 & 0.084 & $141.31^{+14.29}_{-13.21}$\\
        102 & NICER & 2020-04-28 & 00:48:00.243 & 1.446 & $2.34^{+0.05}_{-0.05}$\\
        103 & NICER & 2020-04-28 & 00:48:08.251 & 0.453 & $1.51^{+0.15}_{-0.10}$\\
        104 & NICER & 2020-04-28 & 00:48:23.754 & 3.508 & $0.17^{+0.01}_{-0.01}$\\
        105 & NICER & 2020-04-28 & 00:48:27.020 & 1.129 & $1.55^{+0.07}_{-0.07}$\\
        106 & NICER & 2020-04-28 & 00:48:33.367 & 0.938 & $2.34^{+0.11}_{-0.11}$\\
        107 & Fermi & 2020-04-28 & 00:48:44.824 & 0.382 & $62.36^{+10.73}_{-9.32}$\\
        108 & Fermi & 2020-04-28 & 00:48:49.272 & 0.112 & $730.45^{+41.34}_{-34.91}$\\
        109 & NICER & 2020-04-28 & 00:48:56.199 & 3.788 & $1.17^{+0.03}_{-0.03}$\\
        110 & Fermi & 2020-04-28 & 00:49:00.273 & 0.120 & $49.00^{+10.00}_{-10.25}$\\
        111 & Fermi & 2020-04-28 & 00:49:01.121 & 0.151 & $80.60^{+11.13}_{-10.13}$\\
        112 & Fermi & 2020-04-28 & 00:49:01.936 & 0.181 & $58.23^{+13.09}_{-12.15}$\\
        113 & Fermi & 2020-04-28 & 00:49:06.472 & 0.022 & $75.91^{+21.36}_{-20.00}$\\
        114 & Fermi & 2020-04-28 & 00:49:16.592 & 0.234 & $176.24^{+14.06}_{-11.79}$\\
        115 & Fermi & 2020-04-28 & 00:49:22.392 & 0.078 & $80.51^{+10.64}_{-9.87}$\\
        116 & Fermi & 2020-04-28 & 00:49:27.280 & 0.082 & $43.41^{+11.22}_{-9.39}$\\
        117 & NICER & 2020-04-28 & 00:49:32.702 & 0.821 & $0.45^{+0.04}_{-0.03}$\\
        118 & NICER & 2020-04-28 & 00:49:34.115 & 0.853 & $0.79^{+0.06}_{-0.05}$\\
        119 & NICER & 2020-04-28 & 00:49:35.948 & 0.799 & $3.39^{+0.16}_{-0.15}$\\
        120 & NICER & 2020-04-28 & 00:49:36.957 & 1.281 & $0.17^{+0.02}_{-0.02}$\\
        121 & NICER & 2020-04-28 & 00:49:40.293 & 2.106 & $0.32^{+0.02}_{-0.02}$\\
        122 & NICER & 2020-04-28 & 00:49:42.741 & 1.408 & $0.52^{+0.04}_{-0.04}$\\
        123 & Fermi & 2020-04-28 & 00:49:46.142 & 0.036 & $44.17^{+13.61}_{-12.22}$\\
        124 & Fermi & 2020-04-28 & 00:49:46.680 & 0.150 & $144.13^{+18.53}_{-20.93}$\\
        125 & NICER & 2020-04-28 & 00:49:50.079 & 0.484 & $0.27^{+0.05}_{-0.03}$\\
        126 & NICER & 2020-04-28 & 00:49:52.271 & 2.119 & $0.83^{+0.04}_{-0.04}$\\
        127 & NICER & 2020-04-28 & 00:49:55.078 & 1.197 & $0.44^{+0.03}_{-0.03}$\\
        128 & Fermi & 2020-04-28 & 00:50:01.012 & 0.047 & $140.21^{+20.64}_{-18.51}$\\
        129 & Fermi & 2020-04-28 & 00:50:01.358 & 0.095 & $279.05^{+20.42}_{-25.16}$\\
        130 & NICER & 2020-04-28 & 00:50:03.786 & 0.803 & $1.48^{+0.11}_{-0.10}$\\
        131 & NICER & 2020-04-28 & 00:50:05.664 & 0.440 & $0.72^{+0.07}_{-0.05}$\\
        132 & NICER & 2020-04-28 & 00:50:09.487 & 0.424 & $0.43^{+0.07}_{-0.06}$\\
        133 & NICER & 2020-04-28 & 00:50:11.182 & 0.624 & $0.17^{+0.03}_{-0.02}$\\
        134 & NICER & 2020-04-28 & 00:50:14.097 & 1.952 & $0.29^{+0.02}_{-0.02}$\\
        135 & NICER & 2020-04-28 & 00:50:17.529 & 1.172 & $0.59^{+0.04}_{-0.04}$\\
        136 & Fermi & 2020-04-28 & 00:50:21.969 & 0.019 & $70.53^{+30.53}_{-25.79}$\\
        137 & NICER & 2020-04-28 & 00:50:30.726 & 0.237 & $0.16^{+0.09}_{-0.03}$\\
        138 & NICER & 2020-04-28 & 00:50:34.936 & 0.674 & $1.95^{+0.09}_{-0.09}$\\
        139 & HXMT$^3$ & 2020-05-06 & 22:48:21.550 & 0.035 & $7.60^{+1.40}_{-1.50}$\\
        140 & HXMT & 2020-05-12 & 21:47:43.340 & 0.015 & $10.00^{+2.20}_{-1.90}$\\
        \hline
    \end{tabular}
    \begin{tablenotes}
        \item[$\dagger$]{(10$^{-8}$ erg cm$^{-2}$ s$^{-1}$).}
        \item[*]{Flux data not reported.}
        \item[1]{Fermi detections are provided in the energy range of 8 keV--1 MeV \citep{yang_bursts_2021}.}
        \item[2]{NICER detections are provided in the energy range of 0.5--10 keV \citep{younes_nicer_2020}.}
        \item[3]{HXMT detections are provided in the energy range of 1--250 keV \citep{2022ApJS..260...25C}.}
    \end{tablenotes}
    \end{threeparttable}
\end{table}

\section{Characteristic Spectral Width of FRB 200428 Referenced to Peak Frequency}
\label{appx:spectral_width}

The two independent detections of FRB 200428 by NICER and CHIME can be used to estimate a characteristic spectral width of FRB 200428 assuming it has a Gaussian spectral shape.

From equation~\ref{eq:specific_luminosity_gaussian},
\begin{equation}
    f_{\nu,\text{CHIME}} = f_{\nu,\text{peak}}\exp\left[-\frac{1}{2}\left(\frac{\nu_{\text{CHIME}}-\nu_{\text{peak}}}{\delta\nu}\right)^2 \right]
\end{equation}
where $f_{\nu,\text{peak}}\approx1320$ MHz is the peak frequency of FRB 200428 (Extended Data Fig. 1 in \cite{bochenek_frb_2020}) and $f_{\nu,\text{CHIME}}\approx600$ MHz is the central observing frequency of CHIME. Solving for $\delta\nu$,
\begin{equation}
    \delta\nu = \frac{|\nu_{\text{CHIME}}-\nu_{\text{peak}}|}{2\left[\ln(f_{\nu,\text{peak}})-\ln(f_{\nu,\text{CHIME}})\right]^{1/2}}
\end{equation}
one obtains a characteristic burst spectral width for FRB 200428, $\delta\nu\approx367$ MHz. Referenced to the peak frequency, $\delta\nu/\nu\approx(367\text{ MHz})/(1320\text{ MHz})=0.278$.


\bsp	
\label{lastpage}
\end{document}